\documentclass[12pt,preprint]{aastex}
\usepackage{epsfig,amssymb}
\usepackage{graphicx}
\usepackage{rotating}
\usepackage{comment}
\usepackage{natbib}
\newcommand{\ben}{\begin{enumerate}}
\newcommand{\een}{\end{enumerate}}
\newcommand{\bfig}{\begin{figure}}
\newcommand{\efig}{\end{figure}}
\newcommand{\beq}{\begin{equation}}
\newcommand{\eeq}{\end{equation}}
\newcommand{\mbf}{\mathbf}

\shorttitle{Non-Neutralized Currents and Lorentz Forces in Active Regions}
\shortauthors{Georgoulis, Titov, \& Miki\'{c}}

\begin{document}
\title{Non-neutralized Electric Current Patterns in Solar Active Regions: Origin of the Shear-Generating Lorentz Force} 


\author{Manolis K. Georgoulis\footnote{Marie Curie Fellow.}}
\affil{Research Center for Astronomy and Applied Mathematics of the 
  Academy of Athens\\
4 Soranou Efesiou Street, Athens, GR-11527, Greece}
\and
\author{Viacheslav S. Titov and Zoran Miki\'{c}}
\affil{Predictive Science, Inc., 9990 Mesa Rim Road, San Diego, CA
       92121, USA}
%
%
%
\begin{abstract}
Using solar vector magnetograms of the highest available spatial resolution and signal-to-noise ratio we perform a detailed study of electric current patterns in two solar active regions: a flaring/eruptive, and a flare-quiet one. We aim to determine whether active regions inject  non-neutralized (net) electric currents in the solar atmosphere, responding to a debate initiated nearly two decades ago that remains inconclusive. We find that well-formed, intense magnetic polarity inversion lines (PILs) within active regions are the only photospheric magnetic structures that support significant net current. More intense PILs seem to imply stronger non-neutralized current patterns per polarity. This finding revises previous works that claim frequent injections of intense non-neutralized currents by most active regions appearing in the solar disk but also works that altogether rule out injection of non-neutralized currents. In agreement with previous studies, we also find that magnetically isolated active regions remain globally current-balanced. In addition, we confirm and quantify the preference of a given magnetic polarity to follow a given sense of electric currents, indicating a dominant sense of twist in active regions. This coherence effect is more pronounced in more compact active regions with stronger PILs and must be of sub-photospheric origin. Our results yield a natural explanation of the Lorentz force, invariably generating velocity and magnetic shear along strong PILs, thus setting a physical context for the observed pre-eruption evolution in solar active regions. 
\end{abstract}
\section{Introduction}
In spite of its relative thinness, the solar photosphere plays an 
important role as the transitional layer between the fluid-dominated solar
interior and the magnetically dominated solar atmosphere. Over the
past decades the emergence process of magnetic flux 
through the photosphere has been
thoroughly studied by observers and modelers. An early observational
finding was the filamentary, fibril structure of 
photospheric magnetic flux
strands or tubes (e.g., \citealp{livingston_harvey69, 
howard_stenflo72, stenflo73}) that defies the spatial 
resolution of our best magnetographs to this day. 
This fine, fibril structure of
photospheric magnetic fields is yet to be fully understood and 
properly interpreted \citep{parker79, parker07}. 

Soon after the discovery of the filamentary photospheric magnetic fields, it
was realized that emerging flux tubes must be twisted in order to maintain their 
structural integrity \citep{schuessler79}. Recent
simulations of magnetic flux emergence 
(\citealt{murray_etal06, archontis_hood08} and references therein) 
further showed that magnetic flux
tubes cannot even emerge without some twist. Twist implies the presence of 
field-aligned electric currents, realized by the curl of the measured or simulated photospheric magnetic field vector per Ampere's law. The existence of electric currents automatically implies departure of the magnetic structure from its minimum-energy, current-free (potential) configuration, as potential fields are defined by the gradient of a scalar potential calculated in a finite, bounded, or infinite, partly bounded, volume \citep{schmidt64, sakurai82}.  Non-potentiality in solar magnetic fields is the undisputed underlying cause of solar eruptive activity originating typically in complex regions of interacting opposite-polarity magnetic fields, interfacing by strong polarity inversion lines (PILs). Observed photospheric PILs are typically deformed, shredded, and stretched by strong shear flows and associated magnetic shear, measured as the local angular difference between the observed field vector and the expected potential-field configuration (see, e.g., \citet{wang99} for a review, but also \citet{zirin_wang93}, \citet{tiwari_etal10}, and references therein).

In spite of strong mechanical forces and complexity in the photosphere, twist seems to remain almost unchanged in the corona
\citep{mcclymont_etal97}, where the substantially decreased plasma 
$\beta$-parameter 
allows for a nearly complete spatial filling of the coronal volume
by magnetic field lines. The low-$\beta$ coronal environment further
implies almost purely field-aligned currents, except on thin
current sheets that interface between flux tubes \citep{parker72, parker04}. 
As a result, there is a {\it volume} current in 
flux tubes that is {\it non-neutralized} across any cross-section of the
corona (e.g., \citealt{longcope_welsch00}). The notion of non-neutrality of electric currents implies the presence of a net (nonzero) current for a given magnetic polarity, as defined by previous works \citep[e.g.][]{wilkinson_etal92,wheatland00}.

On the other hand, the presence of 
{\it volume} currents in twisted magnetic fibrils embedded
in a relatively field-free space occupied by fluid plasma, 
that is the assumed case for the
photosphere, necessarily implies the existence of return (skin or
sheath) {\it surface} currents at the boundaries of these 
photospheric fibrils \citep{pizzo86, ding_etal87, longcope_welsch00}.
In addition to a field-aligned twist component, therefore,  photospheric electric current density shows a significant component perpendicular to the magnetic field $\mbf{B}$. This cross-field current density partially depends on $\nabla B \times \mbf{B}$-terms, caused by gradients of the magnetic field strength $B$ \citep{zhang01, georgoulis_etal04}. 
As such, this component {should peak on the
surface of the flux tube where the interface between the magnetized and 
field-free mediums occurs. In the simplest case of isolated flux tubes it can be shown (see \S5) that the Lorentz force due to these cross-field currents is mainly due to magnetic pressure (see, e.g., \citeauthor{jackson62}'s (\citeyear{jackson62}) decomposition of the Lorentz force into pressure and tension terms) and tends to expand the flux tube, hence competing with the twist that keeps it together in conjunction with the external non-magnetized plasma pressure. 

Little doubt exists that photospheric magnetic fields are indeed forced i.e., subjected to significant Lorentz forces \citep{metcalf_etal95, georgoulis_labonte04}. Above the photosphere, however, 
the plasma pressure decreases abruptly forcing flux tubes to expand rapidly to fill nearly the entire coronal volume, thus smoothing out large gradients 
$\nabla B$. Because of this,
return currents cannot reach the corona - they fade when magnetic fields become force-free (i.e., showing only a volume current due to the twist along
the field). Details of this process remain to be determined, but an interesting mechanism relying on torsional Alfv\'{e}n waves has been suggested by \citet{longcope_welsch00}.

While it is generally established that a net electric current exists 
in any coronal cross-section, 
debate pertains over whether {\it photospheric} currents are
neutralized. \citet{parker96a} argued that for {\it isolated} magnetic
fibrils embedded in a field-free volume, regardless of twist, 
the total current {\it must} be neutralized 
across the photospheric cross-section of any given flux
tube. Simply, then, 
the vertical components of the volume and the surface currents cancel
themselves in each fibril. \citet{parker96a} further suggested that 
the common practice of inferring the
vertical electric current density $J_z$ via the differential form of 
Amp\'{e}re's law $(\nabla \times \mbf{B})_z$ using photospheric
magnetic field measurements $\mbf{B}$ bears no physical meaning. 
Thus, the presence of non-vanishing vertical currents in 
several observational cases is,
according to Parker, an artifact caused by the limited spatial resolution of the
observing magnetograph that tends to attribute an artificially continuous 
magnetic field where magnetic field is inherently discontinuous. 

Melrose \citeyearpar{melrose91, melrose95} followed a different approach, by pointing out that observations are more consistent with non-neutralized photospheric current patterns. 
Therefore, currents have to emerge from the solar interior with 
the emergence of magnetic flux. That 
$\nabla \cdot \mbf{J} =0$ implies that currents have to close at the
base of the convection zone (i.e., the tachocline) where magnetic
fields are generated. This has profound implications for flare initiation: while 
Parker's neutralized
currents require in-situ storage and release of energy, Melrose's large-scale 
current paths imply the existence of a strong inductive coupling between the coronal flaring volume and the convection zone. 
\citet{melrose95} described magnetic loops as electric circuits (the
$[\mbf{E},\mbf{J}]$-paradigm) and cited observations that showed an
increase of magnetic shear after solar flares 
\citep{wang_etal94} to
argue that the energy released in flares is replenished by pumping
non-potential energy through electric currents from deep in 
the solar interior. 

\citet{parker96b} responded to Melrose's proposition by arguing that the 
$(\mbf{E},\mbf{J})$-paradigm results in a set of dynamical equations 
that are mathematically intractable. In contrast, the magnetohydrodynamical (MHD) description of solar processes, in terms of a magnetic field $\mbf{B}$ and a velocity field $\mbf{u}$ (the $[\mbf{B},\mbf{u}]$-paradigm) is more natural and straightforward. The $[\mbf{E},\mbf{J}]$-paradigm may hold in laboratory conditions where magnetic fields are generated by current-carrying coils wrapped around metals, but not in stellar conditions where the dynamo-generated magnetic fields are the ones giving rise to electric currents.

A number of studies followed or preceded the debate, all focusing on
its observational aspect, namely, whether there is non-neutralized current in the
active-region photosphere meaning that net electric currents are
injected into the atmosphere. The study of \citet{wilkinson_etal92} was the only one to 
conclude that photospheric currents may be neutralized. All 
other studies \citep{leka_etal96, semel_skumanich98, wheatland00, falconer01}
concluded that the active-region photosphere includes 
{\it current-carrying} magnetic flux tubes, although currents are roughly balanced at active-region scales, that is, non-neutralized currents of opposite senses close onto each other in a given active region. 
The inferred non-neutralized currents are interpreted
as sub-photospheric in origin. Where these currents close, however, is unclear from
observations and only \citet{wheatland00} appeared to openly favor
Melrose's electric circuit analog. The above studies used both the
differential form of Amp\'{e}re's law, that \citet{parker96a} had
criticized for failing to provide a physical current, and the integral
form of Amp\'{e}re's law, i.e., 
\beq
I = {{c} \over {4 \pi}} \int _S (\nabla \times \mbf{B}) \cdot \mbf{\hat{n}} dS = 
    {{c} \over {4 \pi}} \oint _C \mbf{B} \cdot d \mbf{l}\;\;,
\label{amp}
\eeq
where the total current $I$ is obtained either 
over a cross-section $S$ of a flux tube, where $\mbf{\hat{n}}$ is the unit
vector normal to $S$, 
or along a closed curve $C$ bounding the cross-section $S$.

Although the expressions in equation (\ref{amp}) are equivalent  
(Stokes' theorem), in practice numerical errors can give rise to notable  uncertainties. First and foremost, the inference of magnetic field components via inversion of Stokes images and profiles includes uncertainties subject to the specific inversion model (see, e.g., \S1.3 of \citet{borrero_ichimoto11} and references therein). Because of these uncertainties and since different numbers of pixels are being used for the differential and the integral forms of equation (\ref{amp}), these uncertainties are different for the two theoretically equivalent expressions of Amp'{e}re's law. The integral representation of Amp\'{e}re's law is often preferred because it involves fewer magnetogram pixels (i.e., the outline of a magnetic polarity, rather than the entire polarity used in the differential representation). Moreover, the azimuthal $180^o$-ambiguity, inherent in vector magnetic fields measured by the Zeeman effect \citep{harvey69}, includes further uncertainties if incorrectly resolved. Further, magnetic field measurements are sometimes affected by Faraday rotation \citep{leka_etal96, su_etal06} that is another source of uncertainties. When evaluating equation (\ref{amp}), finite differencing introduces truncation errors depending on the various differentiation schemes used, whereas the curve integral requires precise knowledge of the sequence of contiguous points on the bounding contour, regardless of this contour's shape complexity.

The differential form of Amp\'{e}re's law gives rise to an electric current density $\mbf{J} \sim \nabla \times \mbf{B}$, of which only the vertical component $J_z$ can be readily calculated in the photosphere. While calculation gives rise to large $J_z$-values, a valid question is whether these current density patterns are real. 
\citet{leka_etal96} provide several interesting
arguments supporting the realism of these currents. The most important is that in case of flux
emergence, the increase in currents cannot be accounted for by
the action of photospheric motions alone. In addition, patterns of the vertical current density $J_z$ are known to reflect the expected morphology of the photospheric magnetic field (see, e.g., \citet{socas_navarro05} and \citet{balthasar06}, among others). This evidence, however, cannot be conclusive as $J_z \sim (\nabla \times \mbf{B})_z$ should naturally reflect the shape of the field. \citet{mcclymont_etal97}, in their Appendix B, provided further physically-based arguments, namely that (i) the strongest non-neutralized currents occur in sunspots, where \citet{parker96a} admits that the fibril state of the magnetic fields probably breaks down, (ii) inclined, spiral flux tubes cannot be maintained in their observed position without a true field-aligned current (i.e. twist) reflected in $J_z$, and (iii) $J_z$ assumes a characteristic pattern along sheared magnetic polarity inversion lines (PILs), with narrow ribbon-like structures extending along both sides of the PIL. This, they argued, implies that the individual sheared flux strands are indeed blurred due to the insufficient instrumental resolution but, nonetheless, the current patterns are real, albeit smoothed and only partially resolved. An important conceptual step along the same lines was achieved by \citet{semel_skumanich98}, who hinted that currents cannot be neutralized in case of {\it interacting} flux tubes with opposite polarities,that is, when a tight PIL interfaces between them. In this case one cannot argue for {\it isolated} flux fibrils, an apparent prerequisite for Parker's neutralized currents.

Most of the above studies, including Parker's, underlined the
importance of studying vector magnetograms with 
exceptionally high spatial resolution, sufficient to provide a better view of the fibril, discontinuous magnetic fields. 
The recent study of \citet{venkata_tiwari09} uses an ultra-high-resolution
vector magnetogram to conclude the absence of net currents in (isolated)
sunspots, as Parker suggested. We also undertake this task here, but with a focus on (i) not restricting the analysis to isolated magnetic structures, and (ii) applying detailed error propagation to enable safe results and subsequent conclusions. We utilize two vector magnetograms obtained by the
Spectro-Polarimeter of the Solar Optical Telescope 
onboard the {\it Hinode} spacecraft \citep{kosugi_etal07}. As we shall see below, the spatial resolution and noise level of the Hinode magnetograms are unparalleled compared to magnetograms used in the previously cited studies of the 1990s. Therefore, these data are most appropriate and timely for the targeted sensitive calculations. 

The data and preliminary analysis are described in \S2. The methodology for calculating the total electric currents of ``individual'' (given the spatial resolution) flux tubes in outlined in \S3. In \S4 we present our results and in \S5 we attempt a heuristic physical interpretation of them, connecting them with the observed evolution in the photosphere. We summarize our study and conclude in \S6. 
\section{Vector magnetogram data processing}
The Spectro-Polarimeter (SP; \citealt{lites_etal01}) is part of the
Solar Optical Telescope (SOT) package onboard the Hinode spacecraft. The
instrument records the four Stokes polarization signals at the 
Fe {\small I} 6301.5 and 6302.5 \AA$\;$photospheric magnetically sensitive spectral lines by scanning a portion of the solar disk with a spectral sampling of 21.6 $m$\AA. The normal scanning mode
consists of 2048 steps at a nominal pixel spacing of $0.1585\arcsec$
and with a full slit length equal to 1024 pixel lengths 
\citep{lites_etal08}. 

The inversion of the Stokes profiles is achieved via the Advanced
Stokes Polarimeter (ASP) Milne-Eddington inversion code 
(\citet{lites_etal93} and references therein). The inversion returns the
magnetic field components with a $1 \sigma$ sensitivity 
$\sigma B_l = 2.4\;Mx\;cm^{-2}\;(G)$, for the line-of-sight field
component, and $\sigma B_{\rm tr} = 41\;Mx\;cm^{-2}$, for the transverse
field component. The full-resolution pixel size being $0.1585\arcsec$,
tests on quiet-Sun measurements revealed a spatial resolution of 
$\sim 0.3\arcsec$ \citep{lites_etal08}. These can hardly be compared with pixel sizes of the order $1\arcsec$ or lower (spatial resolution of the order $2\arcsec$ or lower) and noise levels $\sigma B_l$, $\sigma B_{\rm tr}$ of the order several tens to 100-200 G, respectively, achieved by magnetographs used in the earlier studies of electric current patterns and their neutrality properties.

The resolution of the azimuthal $180^o$ ambiguity in the selected
Hinode vector magnetograms was achieved via the
non-potential magnetic field calculation (NPFC) method of 
\citet{georgoulis05}, refined as discussed in \citet{metcalf_etal06}. 
The NPFC method is quite insensitive to noise \citep{leka_etal09} and performs
well also in case of discontinuous and partially unresolved magnetic flux bundles 
\citep{georgoulis10}, despite claims by \citet{leka_etal09}. The NPFC
method is reliable enough to be the method of choice for the automatic
disambiguation of vector magnetograms obtained by the Vector
Spectromagnetograph (VSM) of the Synoptic Optical Long-Term
Investigation of the Sun (SOLIS) facility 
\citep{georgoulis_etal08, henney_etal09}. 

\subsection{Hinode/SOT data on NOAA active region 10930}
The first Hinode SOT/SP data set refers to 
NOAA active region (AR) 10930, a well-studied region 
observed on 2006 December 11. Figure
\ref{ar930}a depicts a continuum image of the active-region
photosphere, showing a clearly interacting $\delta$-sunspot complex (i.e., interacting, opposite-polarity pattern including at least one strong PIL). The respective NPFC-disambiguated vector magnetogram is shown in Figure \ref{ar930}b. The scanning of the AR started at 13:53 UT and ended at 15:15 UT. Even from
this single snapshot one notices the extreme 
magnetic shear ($\sim 90^o$) along the
main PIL of the AR, whose area is selected and expanded in the inset
of Figure \ref{ar930}b. Notice also the
significant magnetic flux fragmentation everywhere in
the field of view. Flux strands with cross-section of 
$\sim 1\arcsec - 2\arcsec$ are abundant both close to the sunspots and to the
extended plage areas surrounding them. The typical vertical field
of these strands ranges between several hundred $G$ and $\sim 1.5\;kG$. 
Such fragmented field concentrations were 
first observed by the balloon-borne vector
magnetograph onboard the Flare Genesis Experiment (pixel size
$0.18\arcsec$; spatial resolution $\sim 0.5\arcsec$ - 
\citealt{bernasconi_etal02}), but could only be guessed 
in previous studies (see, for example, \citet{zirin88} for a
collection of sunspot groups with notable fibril structure observed in
H$\alpha$ wavelengths). 
Adjacent to the PIL (inset), one notices flux bundles, aligned with the PIL 
with a width as small as $\lesssim 2\arcsec$. These ``channels'' or ``lanes''
were first reported by \citet{zirin_wang93} and were exemplified by 
\citet{wang_etal08} for this particular active region. The total
field strength of these flux bundles is $\sim 2\;kG$, with a vertical 
component $|B_z| \sim 1.5\;kG$. The vertical field component on the
PIL dropping precipitously to zero, the horizontal field shows an 
amplitude $B_{\rm h} \sim (1.5 - 1.9)\;kG$. 

Clearly, NOAA AR 10930 does not comply with the Hale-Nicholson
polarity law \citep[e.g.][]{zirin88}: the leading/trailing polarity pattern is unclear and the AR axis shows a North-South, rather than an East-West,
orientation. Given the extremely sheared PIL on the other hand, it is
not surprising that the AR, during its passage from the visible disk 
(i.e., between 12/06/06 and 12/18/06) gave an impressive series of
flares (3 X-class, 2 M-class, about a dozen C-class, and $\sim 100$
B-class events). The most intense of these flares, a GOES X3.4, took
place at 02:14 UT on 2006 December 13, $\sim 1.5$ days after the
Hinode observations of Figure \ref{ar930}. 

A map of the vertical electric current density $J_z$ is shown in
Figure \ref{ar930}c, with the $J_z$-solution for the PIL expanded and
shown in the inset. A brief look at the $J_z$-map roughly confirms the
expected fine structure: in the penumbrae of both sunspots and away from
the PIL there are roughly radial structures in agreement with previous
studies (for example, \citet{semel_skumanich98, balthasar06}, and
others). Further, in a secondary PIL north of the
negative-polarity sunspot there are paired upward/downward
$J_z$-patterns along the PIL. Along the AR's main PIL (inset) the strong
shear indeed creates "tongue"-like upward and downward current patterns. 
Finally, the $J_z$-signature of
the extended plage fields is much weaker, noisier, and lacks a coherent
structure, that is evidence of magnetic fields nearly normal to the
photosphere in these areas \citep{martinez_pillet_etal97}. 

Overall, the vertical electric current density solution for NOAA AR 10930 
shows striking fine structure. 
Magnetic flux fragmentation easily reaches down to the
spatial-resolution capability of the observing instrument. 
Therefore, we cannot claim that the observed magnetic
configuration is fully resolved. The observed $J_z$-structure might, in fact, 
be what \citet{mcclymont_etal97} argued for, namely, the result of
multiple unresolved magnetic fibrils that are imperfectly observed as a
single flux tube.

\subsection{Hinode/SOT data on NOAA active region 10940}
The second data set pertains to Hinode's SOT/SP observations of NOAA
AR 10940, observed on 2007 February 2. The scanning of the AR started
at 01:50 UT and ended at 02:53 UT. Figure \ref{ar940}a shows a
continuum photospheric image of the AR. From this image, the AR
consists of a fairly undisturbed sunspot surrounded by extended plage
and a number of smaller spots, or pores. Evidence for a small
emerging flux sub-region exists just west of the main sunspot, where
the continuum image shows dark fibril structures such as those expected
in arch filament systems \citep{bruzek67}. This area is expanded and
shown in the inset. The PIL of the AR is in the same area but is much
weaker than the main PIL of NOAA AR 10930. The NPFC-disambiguated solution
for the heliographic magnetic field vector is shown in Figure
\ref{ar940}b. The horizontal field vector shows the typical radial
arrangement expected for an undisturbed sunspot while some weak shear 
can be seen along the PIL. 

The solution for the vertical current density, shown in Figure
\ref{ar940}c, is also enlightening: the 
sunspot's penumbra generally exhibits radial $J_z$-patterns except close
to the PIL, where these patterns are deformed (inset). 
There, the $J_z$-patterns roughly follow the 
orientation of the dark fibrils of the arch filament system as seen 
in the continuum (Figure \ref{ar940}a). Weak, plage, 
$J_z$-patterns can also be seen, while the small radial pattern in the
westernmost part of the AR indicates that the darkening in the
continuum image and the coherent negative-polarity feature in the
vector-field image correspond to a small sunspot with almost no umbra.
In this example, too, the $J_z$-patterns provide physical
information on the studied AR. As such, they may not be
completely artificial.

The NOAA AR 10940 is much more quiescent compared to NOAA AR 10930 of
Figure \ref{ar930}. During its passage from the visible disk 
($\sim$ 01/28 - 02/09 2007) the AR gave only 3 weak C-class flares and
$\sim 25$ B-class events. At the time of the Hinode observations on
February 2, the AR was almost totally flare-quiet.

We seek the qualitative and quantitative differences between
the electric current patterns in the two ARs. The findings will be
used to determine, using the
highest possible spatial resolution, whether current-injecting
magnetic flux tubes are widespread, rarely occurred, or simply non-existent. 
\section{The analysis method}
To calculate the total electric current in an entire active region or in polarities of it, we take the following steps: 
first, we translate the pixelized photospheric magnetic-flux distribution into
a collection of partitions, each with a relatively large ($\gg 1$) 
number of pixels. 
Second, we apply the integral form of Amp\'{e}re's law [Equation 
(\ref{amp})] to these partitions to calculate their total electric current.
Third, we algebraically sum the total currents of all partitions to calculate the total current of each polarity and of the entire region. 

Alternatively, the total current can be obtained by, first,
deriving the normal component $J_z$ of the electric current density 
from a vector magnetogram and then integrating the $J_z$-distribution over
the area of a given partition with $N$ pixels or the active region as a whole. 
Calculating the total current from the integral Ampere's law does not rely on
the current densities of all $N$ pixels of a studied partition but it 
uses the magnetic field components of only $\sim \sqrt{N}$ pixels - those that
belong to the perimeter of the partition.
Choosing a sufficiently high magnetic-field threshold in the outline of
the partitions we achieve a relatively high signal-to-noise ratio, and hence a
reasonable uncertainty in our calculations. 
\subsection{Magnetic flux partitioning}
Photospheric flux partitioning is achieved following the partition method of
\citet{barnes_etal05}. Partitioning the vertical component $B_z$ of a magnetogram $\mbf{B}$ aims to ``discretize'' the otherwise continuous magnetic flux distribution, by translating it into a series of well-defined, non-overlapping partitions of different polarities with uniquely defined outlines, areas, and flux contents. Numerically, our method of choice is a ``gradient-based'' flux tessellation scheme that uses a downhill-gradient algorithm \citep[e.g.][]{press_etal92} to define photospheric partitions based on the two-dimensional morphological and polarity characteristics of $B_z$. In addition, a simplification rule is applied, merging partitions that interface by saddle points in $B_z$.

If no thresholds were imposed in the course of partitioning, the entire magnetogram area, including active-region and quiet-Sun patches, would be included in an excessive number of different partitions. It is of little meaning, however, to include weak and/or tiny partitions from the highly fragmented Hinode magnetograms, as these partitions are unlikely to contribute strong currents. For this reason, we focus on strong-flux active-region patches by defining relatively high thresholds for the basic parameters of partitioning:
(i) $B_{z_{thres}}=100\;G$ as the minimum $B_z$-value for each pixel of a given partition, (ii) $\Phi _{thres}=10^{19}\;Mx$ as the minimum magnetic flux $\Phi$ of a partition, and (iii) $S_{thres}=40\;pixels$ as the minimum partition area $S$.  
A partition qualifies for further study only if {\it all} three 
thresholds are exceeded; otherwise, it is discarded.  

A slight mean-neighborhood smoothing is also applied to the original
$B_z$-maps in order to reduce an excessive jaggedness of the
partition contours. Once these contours are determined, however, the original,
non-smoothed maps of the horizontal field components are 
used to calculate the total currents. 
\subsection{Calculation of the total electric current and uncertainties}
\subsubsection{For individual magnetic flux partitions within active regions}
Discretizing the integral Amp\'{e}re's law [Equation (\ref{amp})], we obtain
for each qualifying partition $i$ the total current   
\beq
   I_i = {{c \lambda} \over {4 \pi}} \sum _{k=1}^{K_i}  (B_{x_k} \Delta_{x_k} + B_{y_k} \Delta_{y_k})\, ,
\label{damp}
\eeq
where $\lambda$ is the pixel size in physical units and $K_i$ is the number of
pixels along the contour $C_i$ of the partition. 
Each $k$-th pixel of the contour is characterized by a horizontal
magnetic field $\mbf{B}_{{\rm h}_k} = B_{x_k}  \mbf{\hat{x}} + B_{y_k}
\mbf{\hat{y}}$ and a vector $\Delta\mbf{l}_k = \Delta_{x_k} \mbf{\hat{x}} +
\Delta_{y_k} \mbf{\hat{y}}$, whose components assume one of three possible
values ($0$, $+1$, and $-1$) and represent the unit displacement along the
contour.  

A practical issue is the specification of the sequence of contiguous points that comprise the bounding contour $\Delta \mbf{l}$ of a partition. For this we use an ``edge tracker'' algorithm introduced by \citet{georgoulis_etal12}. The algorithm minimizes the length of the curve by iteratively selecting pairs of adjacent contour points. Contour-length minimization is achieved by a simulated annealing method \citep{press_etal92}.

From Equation (\ref{damp}), we can estimate the 
corresponding uncertainty $\delta I_i$ of $I_i$. 
We assume that $\delta I_i$ stems only from the uncertainties 
$\delta B_{x_k} $ and  $\delta B_{y_k}$ of the field components $B_{x_k}$ and $B_{y_k}$,
respectively, while $\Delta_{x_k}$ and 
$\Delta_{y_k}$ are known without uncertainties.
Considering also $\delta B_{x_k}$ and $\delta B_{y_k}$ to be independent from
each other, we obtain from Equation (\ref{damp})  
\beq
  \delta I_i = {{c \lambda} \over {4 \pi}} 
   \sqrt{\sum _{k=1}^{K_i} (\Delta_{x_k}^2 \delta B_{x_k}^2 + \Delta_{y_k}^2 \delta
   B_{y_k}^2)}\, .
\label{err_damp}
\eeq
If $B_{x_k}$ and $B_{y_k}$ are the horizontal heliographic field components on
the image (observer's) plane, then their uncertainties, $\delta B_{x_k}$ and $\delta B_{y_k}$, respectively, can be calculated by implementing the analytical method described in Appendix I of \citet{georgoulis_labonte04}. We use this approach in the following. 

Thus, for each partition $i$ we calculate the total electric
current $I_i$ from Equation (\ref{damp}) and its uncertainty 
$\delta I_i$ from Equation (\ref{err_damp}).
As a sanity check, Equation (\ref{damp}) is also used to
calculate the total electric current $I'_i$ using a 
{\it potential} magnetic field vector calculated from the
disambiguated solution of the photospheric $B_z$-distribution.
An exact potential field implies that $I'_i \equiv 0$ for every $i$.
However, due to various numerical effects, all $I'_i$ show a nonzero value. 
Comparison between $I_i$ and $I'_i$ can help assess the significance of the nonzero total current $I_i$ of a given partition $i$ and hence the non-neutrality of this partition.

To calculate the photospheric potential field $\mbf{B_p}$ from a given
photospheric $B_z$  distribution, we apply the analytical Green's function method
proposed by \citet{schmidt64}. 
This method is much slower, but quite more accurate - for a given choice of boundary conditions - than the numerical fast-Fourier-transform method of \citet{alissandrakis81}. 
In the analytical method we assume $\mbf{B_p}= - \nabla \psi$, where $\psi$ is
a smooth scalar potential, i.e.,  
\beq
  \psi (\mbf{r}) = {{1} \over {2 \pi}} \int \int 
  {{B_z(\mbf{r'})} \over {|\mbf{r} - \mbf{r'}|}} dx' dy'\;\;,
\label{Bp}
\eeq
defined at a given point $\mbf{r}=x \mbf{\hat{x}} + y \mbf{\hat{y}}$ of 
the photospheric plane. $B_z$, on the other hand, 
is a known function of position 
$\mbf{r'}=x' \mbf{\hat{x}} + y' \mbf{\hat{y}}$ $(\mbf{r} \ne \mbf{r'})$. 

Taking into account the above uncertainties, we characterize a given partition
$i$ as {\it non-neutralized} if 
\beq
  |I_i| > 5\, |I'_i| \mbox{\ and \ }
  |I_i| > 3\, \delta I_i\, .
\label{crit}
\eeq
In other words, we impose a $5 \sigma$ significance level toward the total
currents $I'_i$ of the potential field {\it and} a $3 \sigma$ significance
level toward the calculated uncertainties $\delta I_i$. 

\subsubsection{For active regions as a whole}
In an given active region, let
$N$ be the number of qualifying partitions per the criteria of \S3.1, and  $\Phi_{i}$, $I_i$ be the magnetic flux and total current, respectively, of a given partition $i$. The total currents $I_{+}$ and $I_{-}$ of the positive and negative polarities, respectively, of the region are then given by 
\citep[e.g.][]{wheatland00}
\begin{eqnarray}
\begin{array}{cccccc}
  I_{+} & = & \sum\limits_{i=1}^{N} s^{+}_{i} I_i \, , &  & \mbox{\ where\ } &
  s^{+}_{i} =\frac{1}{2}[1+ {{\Phi _i} \over {| \Phi _i|}} ] \,,\\
  I_{-}  & = & \sum\limits_{i=1}^{N} s^{-}_{i} I_i \, , &   & \mbox{\ where\ }
  &  s^{-}_{i} =\frac{1}{2}[1- {{\Phi _i} \over {| \Phi _i|}} ]\,, 
\end{array}
\label{ipn}
\end{eqnarray} 
with uncertainties 
\begin{eqnarray}
  \delta I_{+} & = & \left( {\sum\limits_{i=1}^N s^{+}_{i} \delta I_i^2} \right)^{1/2} \, ,\\
  \delta I_{-} & = & \left( {\sum\limits_{i=1}^N  s^{-}_{i} \delta I_i^2} \right)^{1/2}\, .
\label{err_ipn}
\end{eqnarray} 
Evidently, significant (i.e., sufficiently larger than applicable uncertainties) $I_{+}$, $I_{-}$ imply non-neutralized currents for the respective polarities. Moreover, the net current in the AR is given by 
\beq
  I_{\rm net} = \sum _{i=1}^N I_i = I_{+} + I_{-}
\label{inet}
\eeq
and has an uncertainty
\beq
\delta I_{\rm net} = \left( { \sum _{i=1}^N \delta I_i^2 } \right)^{1/2} 
= \sqrt{\delta I_{+}^2 +  \delta I_{-}^2}\;\;.
\label{err_inet}
\eeq

The solar plasma is considered 
electrically neutralized, since all the physical processes
in the Sun are relatively slow. 
Then, the conservation of the electric charge in the solar atmosphere
implies that the electric current density $\mbf{J}$ must be solenoidal,
i.e. $\nabla \cdot \mbf{J} = 0$. 
In integral form, this means that the net current through any closed surface,
such as the photosphere, must be zero. 
Taking into account that the strongest coronal magnetic fields and so the
currents are localized within solar ARs, we expect that magnetically isolated ARs should be current-balanced ($I_{net} \sim 0$). In practice, however,
there may be large-scale magnetic
connections between the AR and its surroundings, so the net current may be different from zero. 
This may also happen if the observed area does not include the entire AR.

To jointly characterize imbalances of the total current and the magnetic flux in a
given AR, we introduce the following dimensionless parameters: the 
electric-current imbalance and the magnetic-flux imbalance, 
$\mathcal{I}_{imb}$ and $\mathcal{F}_{imb}$, respectively, where  
\begin{eqnarray}
  \mathcal{I}_{\rm imb} & = & {{|I_{net}|} \over { \sum\limits_{i=1}^N |I_i|}} \, ,
\label{iimb}  \\
  \mathcal{F}_{\rm imb} & = & {{|\Phi _{+} + \Phi _{-}|} \over {\Phi _{+} - \Phi _{-}}}\, . 
\label{fimb}
\end{eqnarray}
Here $\Phi _{+}=\sum\limits^{N}_{i=1}s^{+}_{i} \Phi_{i}$ and $\Phi
_{-}=\sum\limits^{N}_{i=1}s^{-}_{i} \Phi_{i}$ are the total
magnetic fluxes in the positive and negative polarities of the AR,
respectively. 
We will consider the currents in a given AR to be balanced if the inequality
\beq
    \mathcal{I}_{\rm imb}  \ll  \mathcal{F}_{\rm imb}
\label{con2}
\eeq
is satisfied.
This means that the smaller the ratio 
$\mathcal{I}_{\rm imb}/\mathcal{F} _{\rm  imb}$ 
(ideally $\mathcal{I}_{\rm imb} \ll \mathcal{F} _{\rm  imb} \ll 1$), 
the more confident we are that the AR is current-balanced. 

If the criterion of Equation (\ref{con2}) is satisfied, it is also meaningful
to obtain a uniform, normalized degree of neutrality (or non-neutrality, thereof) within
{\it each} polarity of the AR. For this purpose we introduce the 
global (current) non-neutrality factor 
\beq
  \mathcal{I}^{\pm}_{\rm nn} = {{1} \over {2}} ( 
  {{|I_{+}|} \over {\sum _{i=1}^N |s_i^{+} I_i|}} + 
  {{|I_{-}|} \over {\sum _{i=1}^N |s_i^{-} I_i|}} )\;\;.
\label{ipmimb}
\eeq
This dimensionless 
parameter determines the degree to which different partitions of the
same polarity show the same sense of total electric current. 
For strongly non-neutralized, but also {\it coherent} polarities, whose partitions predominantly show a given sense of non-neutralized total current, one obtains $\mathcal{I}_{\rm imb} \ll \mathcal{I}^{\pm}_{\rm nn} \sim 1$.
For very incoherent, or current-neutralized, polarities, 
$\mathcal{I}_{\rm imb} \sim \mathcal{I}^{\pm}_{\rm nn} \ll 1$. The
intermediate cases $\mathcal{I}_{\rm imb} \ll \mathcal{I}^{\pm}_{\rm nn} < 1$
and $\mathcal{I}_{\rm imb} < \mathcal{I}^{\pm}_{\rm nn} < 1$ imply a degree of
incoherence and/or non-neutrality for each polarity. 
\section{Current patterns in the selected ARs}
\subsection{NOAA AR 10930}
Figure \ref{ar930p}a shows the flux-partitioned vertical magnetic
field of NOAA AR 10930. We identified a total of $N=531$ flux
partitions, of which $p=246$ show positive polarity and $n=285$ show
negative polarity. To cross-check our calculation, 
we infer the mean vertical electric current density of each
partition in two different ways: first, by dividing the total current
$I_i$ of each partition $i$ by the area $S_i$ of the partition 
($J_{z_i} = I_i / S_i$) and, second, by averaging all the local
vertical current densities $J_{z_q}$ within each partition $i$ 
($J_{z_i} = \langle J_{z_q} \rangle$), where $J_{z_q}$ have been
obtained from the differential expression of Amp\'{e}re's law. Close correlation between these two $J_z$-maps implies precise calculation of the bounding contours of the partitions. The results from the two mean-$J_z$ calculations are depicted in Figures  \ref{ar930p}c and  \ref{ar930p}d, respectively, and are
compared in Figure \ref{ar930p}b. Evidently there is close agreement
between the two alternative current density calculations. 
The generally small discrepancies are mostly caused by uncertainties in the
calculation of the differential $J_{z_q}$ via finite differencing. This
action is expected to slightly overestimate the 
magnitudes $|J_{z_i}|$ of the 
differential $J_{z_i}$. Indeed, Figure \ref{ar930p}b shows 
$(J_z)_{integral} \sim 0.94 (J_z)_{differential}$, 
implying a flatter least-squares best 
fit (black line) than the analytical equality (red line).  

We now focus on the total
electric currents of individual partitions. In random order, these
currents appear in Figure \ref{ar930cl}a. We also check which of these
currents survive our significance test of equations
(\ref{crit}). Notably, 120 of the 531 identified partitions
satisfy the significance criteria. 
Of them, however, only $\sim 25$ have strong enough total
currents to be discernible in Figure \ref{ar930cl}a 
(non-neutralized currents shown with red 
columns and green error bars). Regardless, it is
evident that {\it there are} flux partitions in the AR 
with non-neutralized currents. What is more instructive is the spatial
distribution of these partitions, shown in the detail of Figure \ref{ar930cl}b. We 
find that partitions with the strongest non-neutralized currents
are {\it adjacent} to the PILs of the AR. The two main interacting sunspots
(labeled 1 and 2 in Figures \ref{ar930cl}a and \ref{ar930cl}b) have
large total currents with values $\sim 45 \times 10^{11}\;A$ and 
$\sim -35 \times 10^{11}\;A$, respectively. Partitions adjacent to the secondary
PILs of the AR, some of them labeled (3-5), have total currents of the
order $(5-10) \times 10^{11}\;A$. There is virtually no partition with
clearly non-neutralized total current that is far from the AR's PILs. We
conclude, therefore, that the intensely flaring NOAA AR 10930 includes
strong non-neutralized currents {\it only} where 
opposite-polarity flux-tube footprints are close enough to 
interact, that is, in proximity to PILs. 

NOAA AR 10930 as a whole has a net current 
$I_{\rm net} = (8 \pm 1.6) \times 10^{11}\;A$. With much stronger $I_{+}$
and $I_{-}$ ($\pm [80-90] \times 10^{11}\;A $), 
the AR has a total current
imbalance $\mathcal{I}_{\rm imb} \simeq 0.036$. This is more than an order of
magnitude smaller than the magnetic flux imbalance of the AR, 
$\mathcal{F} _{\rm imb} \simeq 0.4$. Consequently, NOAA AR 10930 
as a whole can be considered current-balanced. This being said,the
non-neutrality factor $\mathcal{I}^{\pm}_{\rm nn} \simeq 0.80$, so 
$\mathcal{I}^{\pm}_{\rm nn} \gg \mathcal{I}_{\rm imb}$. As a result, besides
strong non-neutrality in the AR, there is a large degree of coherence
in the sense of electric current for each polarity: in particular, 
$\sim 80$\% of partitions of 
a given polarity show a given sense of electric current 
(positive/negative current for negative/positive magnetic polarities). 
This opposite current-sense preference per polarity achieves an overall current-balanced AR so that non-neutralized currents of opposite polarities approximately close onto each other. 
\subsection{NOAA AR 10940}
The flux-partitioned vertical magnetic field of NOAA AR 10940 is
depicted in Figure \ref{ar940p}a. Here we identified a total of
$N=297$ partitions, of which $p=156$ show positive polarity and
$n=141$ show negative polarity. The average vertical current densities
for each partition via the integral and the differential
calculation are shown in Figures \ref{ar940p}c and \ref{ar940p}d,
respectively, and
are compared in Figure \ref{ar940p}b. For this flare-quiet AR the
average current densities are generally smaller than those of the flaring
NOAA AR 10930. As expected, the differential $J_z$ slightly overestimates the electric current density. From Figure \ref{ar940p}b, $(J_z)_{integral} \sim 0.93 (J_z)_{differential}$ (black line). 

Important quantitative differences 
between NOAA ARs 10930 and 10940 are revealed, however, when the
total currents of partitions are plotted for NOAA AR 10940: Figure \ref{ar940cl}a
provides these currents in random order. The peak partition
current here barely exceeds $3 \times 10^{11}\;A$ that is more than an order
of magnitude smaller than the respective peaks in
NOAA AR 10930 (Figure \ref{ar930cl}a). Of the 297 identified
partitions, 51 survive the significance test of equations
(\ref{crit}). Of them, only 10 have total currents in excess of
$10^{11}\;A$. These partitions are all close to the weak, loosely formed
PIL of the AR (Figure \ref{ar940cl}b). We conclude therefore, that 
(i) the flare-quiet NOAA AR 10940 also includes non-neutralized electric currents, albeit much weaker than those of the flaring NOAA AR 10930, and (ii) these weaker, non-neutralized currents are also found exclusively along the AR's weaker PIL. It appears that the stronger
(more flux-massive) the PIL, the more intense the non-neutralized currents
it involves. 

NOAA AR 10940 as a whole has a net current 
$I_{\rm net} = (3.8 \pm 1.7) \times 10^{11}\;A$, comparable 
to that of NOAA AR 10930, although the total currents $I_{+}$
and $I_{-}$ are substantially smaller here 
($\pm [10-15] \times 10^{11}\;A$). 
The total current imbalance in the AR is $\mathcal{I}_{\rm imb} \simeq 0.063$, 
$\sim 4.5$ times smaller than the magnetic flux imbalance 
$\mathcal{F} _{\rm imb} \simeq 0.28$. Moreover, the non-neutrality factor 
$\mathcal{I}^{\pm}_{\rm nn} \simeq 0.40$, so  
$\mathcal{I}^{\pm}_{\rm nn} > \mathcal{I}_{\rm imb}$. This degree
of coherence in NOAA AR 10940 is half that of the flaring NOAA
AR 10930: different polarities show a significantly weaker preference in a
given sense of electric currents. 

Summarizing our findings, 
(i) NOAA AR 10940 as a whole can be characterized
current-balanced, just like NOAA AR 10930, 
(ii) NOAA AR 10940 includes
much weaker total currents than NOAA AR 10930, 
(iii) {\it both} NOAA ARs 10940
and 10930 involve non-neutralized current patterns flowing 
{\it exclusively} along their PILs, with the ones for the flare-quiet 
AR 10940 (that has a much weaker PIL) being more than an order of magnitude 
weaker than those of the flaring AR 10930, and 
(iv) NOAA AR 10940 is much less coherent (a factor-of-two difference) in terms of preference of electric-current sense per polarity, compared to NOAA AR 10930. 
\subsection{Dependence of results on partitioning threshold and varying spatial resolution}
To obtain the results of \S\S 4.1 and 4.2 we used a fixed, 
arbitrarily chosen threshold 
$B_{z_{thres}}=100\;G$ for the magnetic-flux partitioning and we
took advantage of the high spatial resolution of Hinode's SOT/SP 
magnetograms. To complete the analysis, we investigate the role
of both the arbitrary $B_{z_{thres}}$ and the spatial resolution in
our findings. 

\subsubsection{Varying partitioning threshold}
First, we vary the value of $B_{z_{thres}}$ from $100\;G$ to
$1000\;G$ and perform the same analysis as in \S\S 4.1 and
4.2. We find that non-neutralized currents in both ARs
continue to accumulate in the partitions identified in Figures
\ref{ar930cl} and \ref{ar940cl}, that is, close to the PILs of the
ARs. In Figure \ref{ar930_stat} we show this behavior for NOAA AR 10930. 
In Figure \ref{ar930_stat}a, the black curve indicates the
fraction of the total (unsigned) magnetic flux of the AR that is
represented by the partitioning for each $B_{z_{thres}}$-value. 
Understandably, as more flux is left out of the
partitioning for increasing $B_{z_{thres}}$, this fraction
decreases. The magnetic flux imbalance $\mathcal{F}_{\rm imb}$ and the
total current imbalance $\mathcal{I}_{\rm imb}$ for various $B_{z_{thres}}$ are
shown with blue and red curves, respectively. For all
$B_{z_{thres}}$, $\mathcal{I}_{\rm imb}$ is always 5-40 times smaller than 
$\mathcal{F}_{\rm imb}$, regardless of the value of
$B_{z_{thres}}$. Therefore, irrespectively of $B_{z_{thres}}$,
NOAA AR 10930 includes non-neutralized current patterns close to its
PILs but stays current-balanced globally. The non-neutrality factor 
$\mathcal{I}^{\pm}_{\rm nn}$ is substantial at all times ($\gtrsim 0.8$) 
and increases nearly monotonically for increasing $B_{z_{thres}}$. 
Therefore, we can safely conclude that
$\mathcal{I}_{imb} \ll \mathcal{I}^{\pm}_{\rm nn} \sim 1$ 
so the two different polarities in NOAA AR 10930 have both strong non-neutralized currents and a clear preference of
sense of electric currents, independently of $B_{z_{thres}}$. 

In Figure \ref{ar930_stat}b the blue curves show $I_{+}$, for the
positive-polarity partitions, and $I_{-}$ for the negative-polarity ones,
while the red curve shows the net current $I_{net}$. That $I_{+} <0$ and 
$I_{-} >0$ indicates a predominantly left-handed, counterclockwise twist in
the AR. Notice also that $|I_{+}|$ and $|I_{-}|$ tend to decrease for
increasing $B_{z_{thres}}$. This indicates that non-neutralized currents become weaker if
stronger, less interacting, fields are chosen for partitioning. 
Indeed, moving to thresholds $B_{z_{thres}}$ from several hundred $G$ to $1\;kG$, 
we exclude some parts of the PILs of the AR. Regardless, the total
current balance of the AR does not depend sensitively on $B_{z_{thres}}$:
$|I_{net}|$ is always $\sim 3.5 - 4$ times smaller than $|I_{+}|$ and $|I_{-}|$.

The same analysis on NOAA AR 10940 is depicted in Figure
\ref{ar940_stat}. Because of the much weaker currents involved, increasing 
$B_{z_{thres}}$ in this case leads to more striking quantitative changes: for 
$B_{z_{thres}}$ ranging between $400\;G$ and $800\;G$ the current imbalance 
$\mathcal{I}_{\rm imb}$ increases to $\sim 0.5$ but still
remains smaller than the respective flux imbalance $\mathcal{F}_{\rm imb}$
that ranges between 0.5 and 0.6 for the same $B_{z_{thres}}$-range (Figure
\ref{ar940_stat}a). For $B_{z_{thres}} > 800\;G$, $\mathcal{I}_{\rm imb}$
decreases significantly to reach $\sim 0.2$ at 
$B_{z_{thres}} = 1000\;G$. Apparently the combination of weak total currents
(Figure \ref{ar940_stat}b) and a strongly imbalanced, inhomogeneous magnetic
flux distribution where the one or the other polarity is favored for different 
$B_{z_{thres}}$-values is the reason for these effects. 
The non-neutrality factor 
$\mathcal{I}^{\pm}_{\rm nn}$ initially increases and remains high at $\sim 0.7$
until $B_{z_{thres}} \simeq 600\;G$, thereafter decreasing significantly to
reach a value $\sim 0.4$. At all times, though, 
$\mathcal{I}^{\pm}_{\rm nn} > \mathcal{I}_{\rm imb}$. The 
polarity currents $I_{+}$ and $I_{-}$ 
(Figure \ref{ar940_stat}b) peak at $\sim 20 \times 10^{11}\;A$, $\sim 5$ times
smaller than those in NOAA AR 10930. Their magnitudes also decrease for
increasing $B_{z_{thres}}$ and this causes $I_{net}$, and hence 
$\mathcal{I}_{\rm imb}$, to change significantly. 
Nonetheless, it would be biased
to claim that whether NOAA AR 10940 is globally 
current-balanced depends on the
value of $B_{z_{thres}}$. For reasonable values of this threshold, sufficient
to keep a reasonably balanced representation of both polarities in the
partitioning, the flare-quiet NOAA AR 10940 has an overall 
current-balanced
structure, includes non-neutralized currents along its PIL,  
and shows some preference in a given sense of current for each
polarity, similarly to its flaring counterpart NOAA AR 10930. However, this
preference is much weaker than in the flaring AR, 
as are the non-neutralized current patterns NOAA AR 10940 involves. 

\subsubsection{Varying spatial resolution}
We now study the impact of changing spatial resolution to the current
patterns of the two ARs. We first bin each magnetogram using different 
sampling factors and then we re-calculate the subsequent current patterns using the analysis of \S3 and keeping a
fixed $B_{z_{thres}} =100\;G$ in all cases. For NOAA ARs 10930
and 10940 the results are summarized in Tables \ref{tb1} and
\ref{tb2}, respectively. 

For both ARs, we reach the following conclusions: 
\ben
\item[(1)] Decreasing the spatial resolution causes a decrease in the
  number of the identified partitions and a subsequent decrease in the
  number of partitions that include non-neutralized currents, that is, 
  those satisfying the significance criteria of Equations
  (\ref{crit}). For the coarsest spatial resolution (pixel size 
  $\sim 2.5\arcsec$) both NOAA ARs 10930 and 10940 include only three
  partitions. The identification of fewer partitions for coarser
  spatial resolution occurs because the
  morphological properties of the photospheric magnetograms - where
  partitioning relies - are
  suppressed or smoothed out when the resolution is reduced.  
\item[(2)] Decreasing the spatial resolution leads to a decrease in
  the total current magnitudes $|I_{+}|$ and $|I_{-}|$. Even in this case,
  however, strong non-neutralized current patterns in the flaring NOAA
  AR 10930 are easily discernible. This is not the case for 
  NOAA AR 10940, whose current magnitudes fall almost within 
  the uncertainty margins ($\sim 10^{11}\;A$) for pixel
  sizes equal to $\sim 1.3\arcsec$ and $\sim 2.5\arcsec$. That 
  NOAA AR 10930 seems to largely maintain its strong non-neutralized
  currents around its PILs even for very coarse resolution justifies
  previous studies that advocate for non-neutralized currents with
  much lower-resolution magnetograms 
  \citep{leka_etal96, semel_skumanich98, wheatland00, falconer01}.
  Our results differ from those of \citet{wilkinson_etal92} 
  who did not find compelling evidence of non-neutralized
  current patterns in the flaring NOAA AR 2372. 
 \item[(3)] Regarding the ARs' overall current balance, Table
   \ref{tb1} suggests that NOAA AR 10930 remains 
   current-balanced regardless of spatial resolution. For
   all cases, $\mathcal{F}_{\rm imb}$ is 5 - 25 times larger than
   $\mathcal{I}_{\rm imb}$. NOAA AR 10940 also appears 
   current-balanced
   except in case the pixel size becomes $\ge 1.3\arcsec$. 
   Then, $\mathcal{I}_{imb} \sim \mathcal{F}_{imb}$ but with very weak 
   $I_{+}$ and $I_{-}$. These weak, insignificant currents practically 
   invalidate any conclusions about $\mathcal{I}_{\rm imb}$.
\item[(4)] Regarding the non-neutrality factor 
  $\mathcal{I}^{\pm}_{\rm nn}$, it appears that binning
  consistently {\it increases}
  the coherence of each polarity in terms of sense of
  electric currents. This is the case for both ARs, although the results are
  somewhat mired in case of the flare-quiet NOAA AR 10940, where 
  $\mathcal{I}_{\rm imb} \sim \mathcal{I}^{\pm}_{\rm nn}$ for coarser pixel
  sizes, due to the much weaker - and hence more uncertain - currents involved. 
\een

Briefly, we conclude that varying the partitioning threshold and/or altering the spatial resolution does not affect our findings of (i) non-neutralized current patterns along PILs, and (ii) globally current-balanced ARs. 
Only very restrictive partitioning
or very coarse spatial resolution can affect these findings, and only
for flare-quiet ARs lacking tight, well-organized PILs. Large
partitioning thresholds or severe binning can cause
artifacts, such as the disappearance of these weak PILs. PILs, however, are
parts of the ``topological skeleton'' of ARs 
\citep{bungey_etal96}, which means that 
they are topologically stable features. As such, at
least the strongest of them will
survive, regardless of thresholding or smoothing.  
This study shows that non-neutralized current patterns are
exclusively related to PILs; apparently  
the stronger the PIL, the more intense the
non-neutralized currents it involves. If well-formed 
PILs cannot disappear due to 
thresholding or binning, their currents cannot disappear under these
actions, either. 
\section{Physical implications of our results}
We now attempt to establish a qualitative connection between our findings and
the observed evolution along well-formed PILs. A crucial aspect of this evolution is
magnetic shear. Shear is {\it invariably} formed along PILs 
\citep[e.g.,][]{falconer_etal97,yang_etal04}
with the
effect being more pronounced in stronger PILs. 
Pending confirmation with larger active-region samples, this work also implies that non-neutralized electric currents are also stronger in stronger PILs. A connection between non-neutralized currents and magnetic shear, therefore,
appears plausible. 

Another notable finding of this study is that partitions of the same
polarity tend to have the same sense of non-neutralized currents. This is
reflected in the non-neutrality factor 
$\mathcal{I}^{\pm}_{\rm nn}$ that is $\sim 1$ for the eruptive NOAA AR 10930, despite the large number
(hundreds) of partitions for each polarity. The same result in the same active region was recently reported by \citet{ravindra_etal11}. A similar effect, but at a clearly smaller degree, occurs for the non-eruptive NOAA AR 10940. This coherence of
electric currents, more pronounced in more compact ARs with stronger PILs,
must be of sub-photospheric origin. Indeed, numerous simulations of magnetic
flux emergence,  
\citep[e.g.,][]{tortosa-andreu_moreno-insertis09,archontis_hood10,cheung_etal10}, 
all start from a
single flux tube with a given twist in the convection zone. Upon emergence in
the photosphere, the flux tube undergoes substantial fragmentation. 
We show here that in the course of such a fragmentation electric currents largely 
retain their sub-photospheric sense. Strong PILs likely indicate a strongly
twisted, perhaps braided \citep{lopez-fuentes_etal03}, coherent sub-photospheric flux tube, while weak PILs suggest a
loosely formed sub-photospheric tube with only the necessary coherence to
survive its emergence in the solar atmosphere. 

But how, and why, 
does magnetic shear occur along intense PILs? For NOAA AR 10930,
studied here, \citet{su_etal07} concluded that the observed shear is due to
sunspot rotation and the east-west motion of the emerging positive-polarity
sunspot just south of the negative-polarity main sunspot of the AR. The main
PIL of the AR is between these two spots (Figure \ref{ar930}). Here we reveal
additional information, namely that strong and systematic 
non-neutralized currents are formed along the PIL and only there. 
These currents suggest that the Lorentz force 
may be the most natural cause of shear. This
is already suggested in several works: \citet{manchester_low00}
analytically showed that the {\it tension} component of the Lorentz force can
cause shear in an undulated emerging flux tube. This result was numerically
demonstrated by \citet{manchester_01}. \citet{manchester_etal04} 
modeled a Lorentz-force-driven shear with a gradient ranging from
photosphere to corona, supporting earlier observations of differential shear
in active regions \citep{schmieder_etal96}. A systematic modeling of the
Lorentz-force-driven shear that demonstrates a coupling between
sub-photospheric and atmospheric magnetic fields was performed by 
\citet{manchester_07}. This author demonstrated by MHD
flux emergence simulations that a gradient in the axial magnetic field (along
the PIL) during the emergence of an $\Omega$-loop is the cause of the Lorentz
force that further causes shear flows leading to magnetic shear. These flows
have an amplitude of about half the local Alfv\'{e}n speed, hence increasing
from the photosphere to the corona. These results were supported by \citet{fang_etal10} who further found that $U$-loops may be formed along
PILs, in agreement with \citet{tortosa-andreu_moreno-insertis09}. 
In the sub-photospheric part of $U$-loops, however, they showed that the
Lorentz force can unshear the magnetic field lines, contrary to the situation
in and above the photosphere. 

In an effort to provide a heuristic explanation of the shear-generating Lorentz force $\mbf{F}$ and its physical connection to non-neutralized currents, we write $\mbf{F}$ in terms of magnetic tension and pressure \citep{jackson62}, i.e.,
\beq
\mbf{F} = {{1} \over {4 \pi}} [ (\mbf{B} \cdot \nabla) \mbf{B} - 
                                 {{1} \over {2}} \nabla B^2 ]\;\;.
\label{fl}
\eeq
The azimuthal component $F_\varphi$ of the Lorentz force in cylindrical coordinates $(r,\varphi,z)$ is given by 
\beq
F_\varphi = {{1} \over {4 \pi}} [ B_r {{\partial B_\varphi} \over {\partial r}}+
           {{1} \over {2r}} {{\partial (B_\varphi^2 - B^2)} \over {\partial  
                            \varphi}}+ 
           B_z {{\partial B_\varphi} \over {\partial z}} + 
           {{B_r B_\varphi} \over {r}} ]\;\;.
\label{fl_phi}
\eeq
All terms in Equation (\ref{fl_phi}) are due to magnetic tension, with the exception of the term $-(1/(2r)) \partial B^2 / \partial \varphi$ that is due to magnetic pressure. 

We first assume a pair of undisturbed, twisted flux-tube footprints embedded in field-free, plasma-filled space (Figure \ref{fl_cartoon}a). Obviously there is no strong PIL interfacing between the footprints in this case. We further ignore the (generally nonzero) azimuthal Lorentz force in the flux-tube interior and focus on the outer edges of the tube (dark blue and red rings) where sheath currents reside, and we seek the azimuthal Lorentz force $F_{\varphi}$ acting in this area. There, the two terms of $F_{\varphi}$ depending on the radial field component $B_r$ vanish because the field has to be largely azimuthal, with a weaker vertical component $B_z$, to ensure a divergence-free magnetic field vector. After some algebra, Equation (\ref{fl_phi}) on the footprint edges becomes 
\beq
F_\varphi \simeq {{B_z} \over {4 \pi}} ( 
           -{{1} \over {r}} {{\partial B_z} \over {\partial \varphi}}+ 
                            {{\partial B_\varphi} \over {\partial z}} )\;\;.
\label{fl_phi1}
\eeq
Notice that {\it all} derivatives in Equation (\ref{fl_phi1}) are due to magnetic tension. Azimuthal symmetry in the undisturbed case 
implies $\partial / \partial \varphi \sim 0$ and since $B_z$ is weak and vanishing at the edge of the flux tubes, $F_\varphi \simeq 0$ from Equation (\ref{fl_phi1}) on the interface between the flux tube and the field-free space. On this interface it is the 
{\it radial} component of the Lorentz force that becomes important. This component, driven by magnetic pressure, will force the flux tube's cross-section to increase as much as the outside plasma pressure allows. The formation of roughly azimuthally symmetric - hence fully neutralized within a given polarity - cross-field sheath (return) currents will protect the flux tube from disintegrating. As \citet{parker96a} demonstrated, there are no net currents injected in this situation. 

We now assume a pair of interacting flux-tube footprints that deform due to this interaction, forming a strong PIL between them (Figure \ref{fl_cartoon}b). The field will still have to be predominantly azimuthal on the interface in this case, too, so Equation (\ref{fl_phi1}) can still be used. However, (i) the vertical field component $B_z$ is non-vanishing any more; in fact, it can be quite strong along both sides of the PIL, and (ii) enhanced, PIL-localized azimuthal field gradients build-up in the deformation area, even in this simplified, axially symmetric (with respect to the PIL) case. These gradients preferentially enhance sheath currents along the PIL. These currents cannot be neutralized any more within a given polarity. Evidently, then, $F_\varphi \ne 0$ along the PIL (but not {\it on} it, because $B_z =0$ there) unless the two terms in the parenthesis of Equation (\ref{fl_phi1}) cancel each other. This resulting azimuthal  Lorentz force is purely due to magnetic tension. Cancellation of the derivative terms may conceivably happen, but in general it is not the cause, and it is certainly not true in case of asymmetries in the magnetic flux distribution within a given footprint. Nonzero azimuthal Lorentz force will cause shearing motions along the PIL that will further give rise to magnetic shear, as demonstrated in numerical simulations. Qualitatively, this result is similar for all {\it interacting} opposite-polarity flux tubes. 

In the actual solar photosphere, the plasma $\beta$-parameter is high enough for MHD to compete with the hydrodynamics of the non-magnetized plasma. This dense plasma, therefore, should exhibit an {\it inertia} to the action of the Lorentz force that can only be overcome locally, where strong fields give rise to powerful azimuthal Lorentz forces (Equation (\ref{fl_phi1})). An intuitive PIL field strength that, if exceeded, should allow $F_{\varphi}$ to move the plasma, is the {\it equipartition} value $B_{eq}$ needed for $\beta =1$. Assuming typical photospheric values for convective flows and mass density, or for number density and the effective temperature, one finds $B_{eq} \sim (200 - 1400)\;G$, with a mean $B_{eq} \simeq 800$ G. In NOAA AR 10930 the field strength around the PIL always exceeds 1.5 kG and, in many cases, it is higher than 2 kG (Figure \ref{ar930}). In this case it is clear that $\beta <1$ in the PIL area, hence $F_{\varphi}$ should be powerful enough to move the plasma. In NOAA AR 10940 the PIL field strength ranges between a few hundred G to $\lesssim$1.5 kG. In this case, therefore, despite some non-neutralized currents, the azimuthal Lorentz force may not be able to shear plasma and magnetic field considerably. This rough interpretation qualitatively agrees with the observational facts for both studied ARs. 

In a symmetric case such as the one of Figure \ref{fl_cartoon}b, Equation (\ref{fl_phi1}) clearly implies an {\it opposite} orientation of $F_{\varphi}$ in the two sides of the PIL that leads to an opposite sense of shear flows. In case of strong asymmetries, however, the competition of the two tension terms in the parenthesis may result in the {\it same} sign of $F_{\varphi}$ in the two PIL sides, and hence in a {\it similar} sense of shear flows. Asymmetries can be simply thought of as severe flux imbalance and/or different field morphologies of the partitions deformed along the PIL. For the studied NOAA AR 10930, for example, inspection of continuum and magnetogram movies implies that velocity shear mainly acts in the counterclockwise direction in {\it both} sides of the PIL. Figure \ref{ar930}b shows clearly that the PIL area is strongly asymmetric. Even in case of similar shear orientation, however, asymmetries lead to strong velocity gradients across the PIL that give rise to strong relative motions in the two sides of the PIL. Albeit in a more complicated manner, therefore, the end result of strongly asymmetric shear flows is qualitatively similar with that of the symmetric case. The directionality of the shear is determined by the direction of $B_\varphi$ that is due to the dominant sense of twist in the flux tube. The  consistency of the shear-flow orientation for the lifetime of the PIL corroborates our findings that there indeed exists a dominant sense of twist in interacting opposite-polarity flux tubes. 

The proposed mechanism of Lorentz-force-driven shear along PILs is, of course, qualitative - careful quantitative studies are necessary for all aspects of it to be clarified at sufficient detail. The study of \citet{torok_kliem03} first showed numerically that as soon as shearing motions are generated in the PIL area, non-neutralized currents inevitably build up in the corona. Shear in that study was generated by placing bipolar flux concentrations -- twisted by photospheric vortex flows -- progressively closer to each other: the closer the concentrations, the larger their non-neutralized total current. Further, \citet{torok_etal11} recently analyzed the evolution of electric currents in an emerging flux-rope simulation. As predicted earlier by \citet{longcope_welsch00}, these authors also found that the electric current in the coronal part of the emerging flux rope is, indeed, essentially non-neutralized, although the physical mechanism underlying this effect has yet to be clarified. Works like the above  substantiate our findings for non-neutralized currents in active-region PILs while our proposed mechanism, at work in case opposite-polarity photospheric flux concentrations approach close enough to interact, provides a physical context for this effect. 
\section{Summary and Conclusions}
Using solar magnetic field measurements of the highest available spatial resolution we have studied in detail the electric current patterns of the photospheric magnetic configurations in two solar active regions; a flaring/eruptive, and a flare-quiet one. Our main objective was to determine whether emerging and/or evolving active regions inject significant  non-neutralized electric currents in the solar atmosphere through the photospheric boundary. We show that such currents are injected solely and exclusively along the photospheric magnetic PILs that accompany the dynamical formation and evolution of active regions. In our limited sample of two active regions we find that stronger PILs imply more intense non-neutralized currents. This result is robust and insensitive to magnetic-field thresholding and the spatial resolution of the studied magnetograms, but has to be confirmed by future studies with larger active-region samples. 

Despite current non-neutrality in particular locations within active regions, our study confirms that active regions as a whole are current-balanced magnetic structures. This means that non-neutralized currents injected on one side of a PIL are roughly counter-balanced by non-neutralized currents of the opposite sense that are developed on the other side of the PIL. 

An additional finding, in agreement with recent results \citep{ravindra_etal11}, is the coherence in the sense of electric currents within a given polarity. We have quantified the effect by means of a dimensionless parameter dubbed the global non-neutrality factor (Equation (\ref{ipmimb})). We find that this effect is more pronounced in the eruptive active region we study, but it is also present in the non-eruptive region (Figures \ref{ar930cl}a; \ref{ar940cl}a and Tables \ref{tb1}, \ref{tb2}), despite the considerable flux fragmentation in the photosphere. We conclude that this effect is of sub-surface origin, which stands in agreement with the results of multiple numerical simulations that show substantial fragmentation of a {\it single} buoyant flux tube in realistic simulations of emerging active regions \citep[e.g.,][]{cheung_etal10}.

Our findings seem to put a nearly twenty-year-old debate to rest: amidst claims ranging between no injection of net electric currents in the solar atmosphere, per \citeauthor{parker96a}'s (\citeyear{parker96a}) isolated flux-tube picture (see Introduction), and injection of net currents in each and every magnetic flux emergence episode, we demonstrate that injection of significant net currents {\it does occur} but it is as rare -- or as frequent -- as the appearance of intense PILs in the solar photosphere. In this case only, Parker's assumption of isolated flux tubes breaks down and \citeauthor{melrose91}'s (\citeyear{melrose91}, \citeyear{melrose95}) proposition of non-neutralized photospheric currents becomes valid. It is obviously beyond the scope of this work to argue on the detailed nature of the sub-photospheric origin of these currents (i.e., the $[\mbf{E},\mbf{J}]$- vs. the $[\mbf{B},\mbf{u}]$-paradigm). However, we argue that strong non-neutralized currents occur exclusively in case of emerging magnetic footprints of opposite polarities that are close enough to interact, i.e. to ``feel'' each other and deform as a result of this interaction. Deformation implies preferential enhancement of return currents sheathing these flux tubes at the deformed areas. Because sheath currents are perpendicular to the axes of the tubes, these perturbations imply the exertion of a Lorentz force that, via magnetic tension, generates shear along the PIL, in terms of both shear flows and magnetic shear. The more enhanced the sheath currents developed in more flux-massive, tight PILs, the stronger the Lorentz force and the resulting shear, provided that the magnetic field strength in the PIL area is greater than the equipartition value (plasma $\beta <1$), so that the Lorentz force can overcome the hydrodynamic inertia and move the plasma (\S5).

The direction of shear flows is typically opposite in the two sides of the PIL (Equation (\ref{fl_phi1})). The magnetic shear angle reflects the sense of twist of the interacting flux tubes: for the studied NOAA AR 10930, negative magnetic polarities associate to positive (non-neutralized) currents and vice-versa (Figures \ref{ar930p}, \ref{ar930cl}) -- this implies a left-handed sense of twist in the region. An opposite situation (right-handed twist) in a similar  field configuration would imply an opposite shear orientation in the region. Notice that the main PIL of NOAA AR 10930 is also strongly asymmetric, leading to predominant counterclockwise, but similarly asymmetric, shear motions on both sides of the PIL. We explain this effect in \S5. These asymmetric flows roughly lead to the same end result as in the case of oppositely oriented shear flows due to significant relative motions in the two sides of the PIL. 

We conclude by emphasizing the need for more detailed qualitative and quantitative studies of the causal sequence {\it interacting flux tubes [PIL]} -- {\it non-neutralized electric currents} -- {\it Lorentz force} -- {\it shear}. Such studies may shed new light into the physics of eruption initiation and pre-eruption active-region evolution. Besides simulations, the role of the Lorentz force appears to become  evident even from observations, albeit in post-eruption situations \citep{hudson_etal08,fisher_etal12}. Moreover, the coherence of non-neutralized currents per polarity, stemming from a common sub-photospheric origin of these currents, directly relates to a {\it dominant} sense of twist in eruptive active regions. This, in addition, may translate to a dominant sense of magnetic helicity in eruptive active regions, as demonstrated by previous  \citep[e.g.,][]{nindos_andrews04,labonte_etal07,georgoulis_etal09} and recent \citep[e.g.,][]{tziotziou_etal12} works.
Undoubtedly, much remain to be revealed, but efforts should focus on putting together as many pieces of the solar eruption puzzle as possible in the most meaningful physical interpretation possible. We intend to undertake such efforts in future studies.   

\acknowledgments
We are grateful to the anonymous referee for critical comments that have soundly improved the presentation of the analysis and results. 
MKG acknowledges the help of Manolis Zoulias in technical matters related to the preparation of the manuscript. This work has received partial support from the European Union's Seventh Framework Programme (FP7/2007-2013) under grant agreement $n^o$ PIRG07-GA-2010-268245. The contributions of VST and ZM were supported by NASA's Living With a Star, Heliophysics Theory, and SR\&T Programs.


\bibliography{refs}
\bibliographystyle{apj}

%
\clearpage
\centerline{\includegraphics[width=12.5cm,height=18.5cm,angle=0]{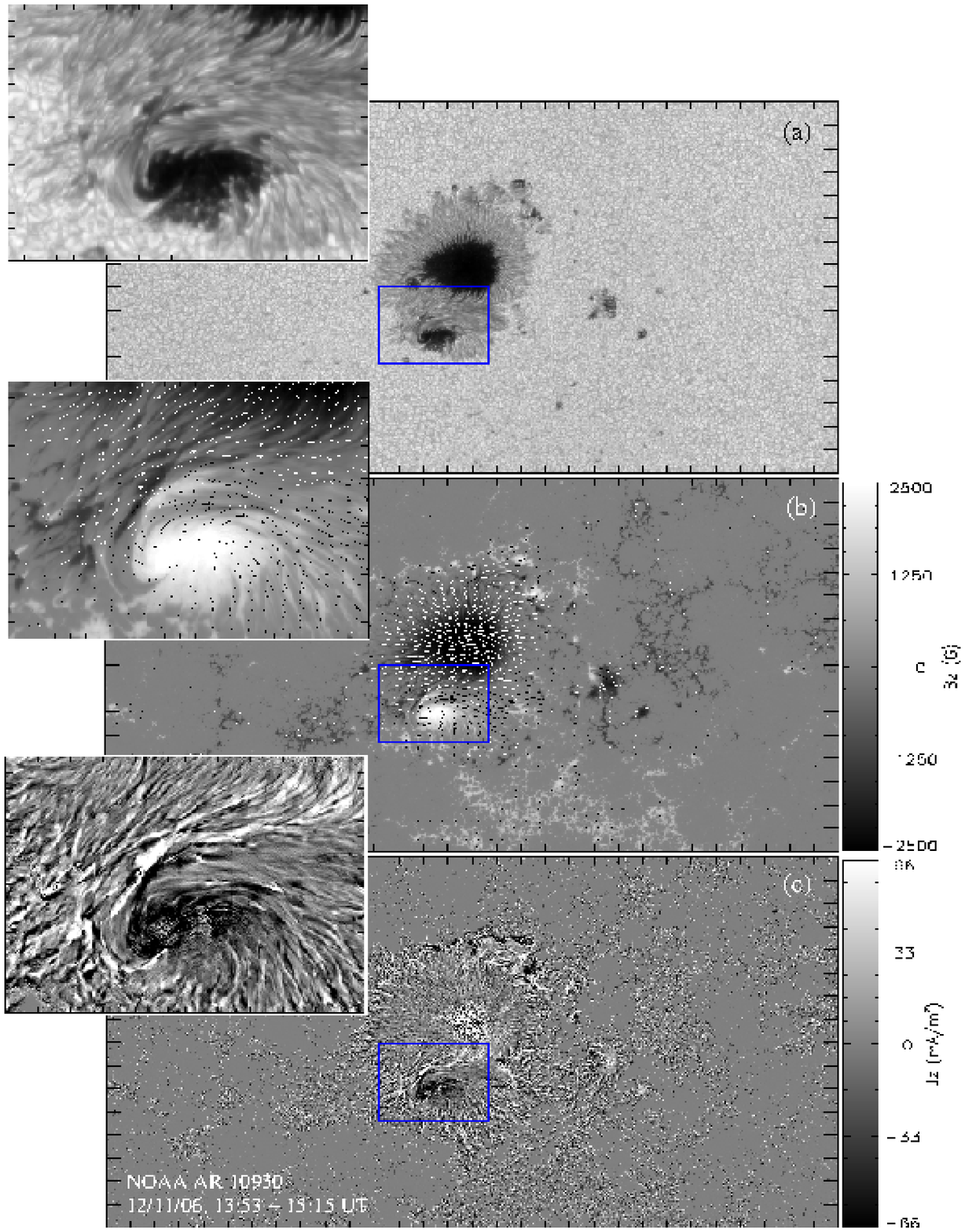}}
\figcaption{
  Views of NOAA AR 10930, observed by SOT/SP on 2006
  December 12, at around 20:30 UT. (a) Continuum image of the
  active-region photosphere. (b) The corresponding photospheric
  magnetic field vector on the heliographic plane. A vector 
  length equal to the tic mark separation corresponds to a horizontal
  field strength of $2000\;G$. (c) The corresponding photospheric
  vertical electric current density. In all images, a part of the AR,
  enclosed by the blue boxes, has been expanded and shown in the
  insets. Tic mark separation for the full images is $10\arcsec$. Tic
  mark separation for the insets is $2\arcsec$. A vector length equal
  to the tic mark separation in the inset of b corresponds to a
  horizontal field strength of $1000\;G$. 
  North is up; west is to the right.
\label{ar930}}
%
\newpage
%
%
\centerline{\includegraphics[width=12.5cm,height=18.5cm,angle=0]{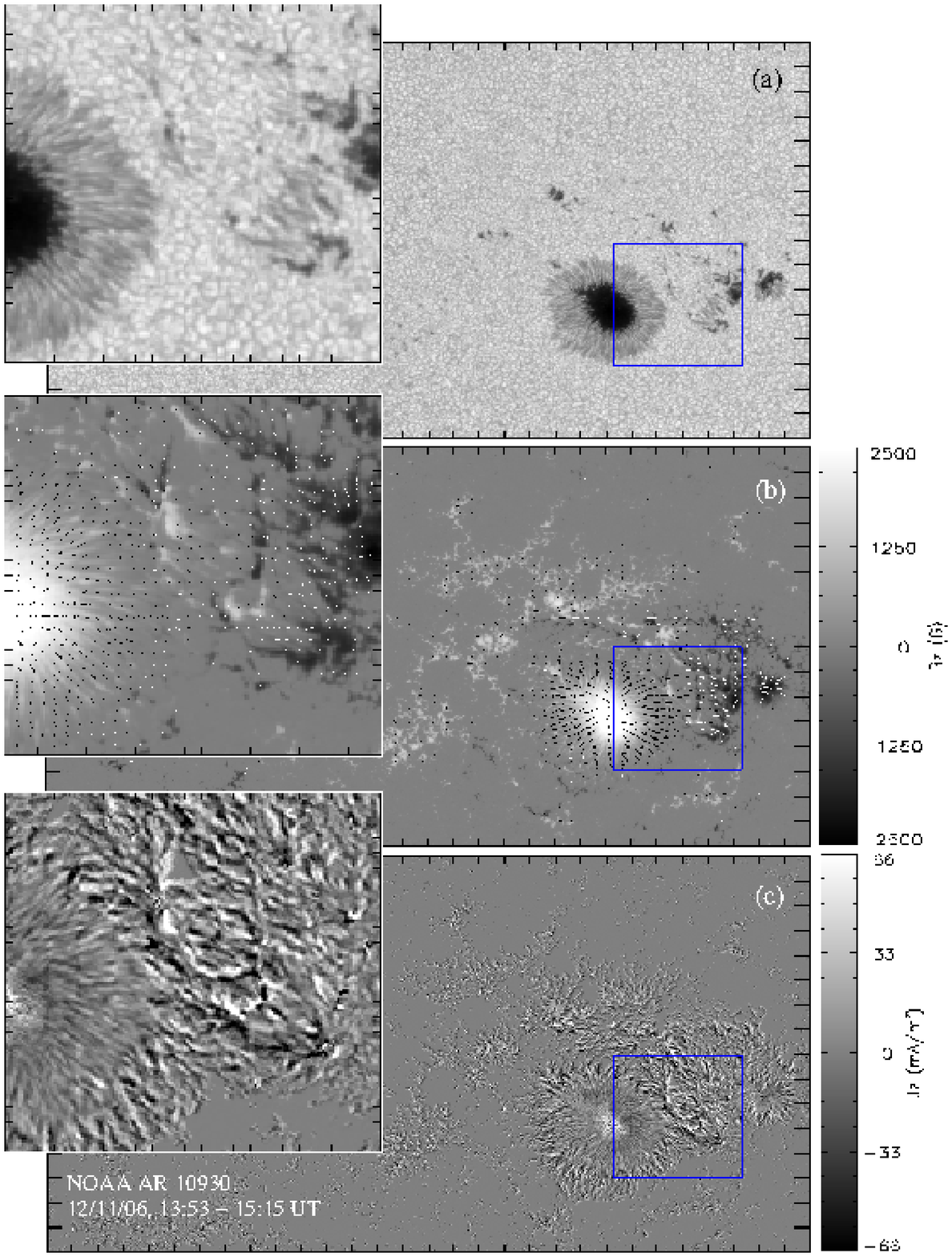}}
\figcaption{Same as Figure \ref{ar930}, but for NOAA AR 10940,
  observed by SOT/SP on 2007 February 2, around 01:50 UT.
  A vector length equal
  to the tic mark separation in the inset of b corresponds to a
  horizontal field strength of $1900\;G$. North is up; west is to the right.
\label{ar940}}
\newpage
\centerline{\includegraphics[width=16.5cm,height=10.cm,angle=0]{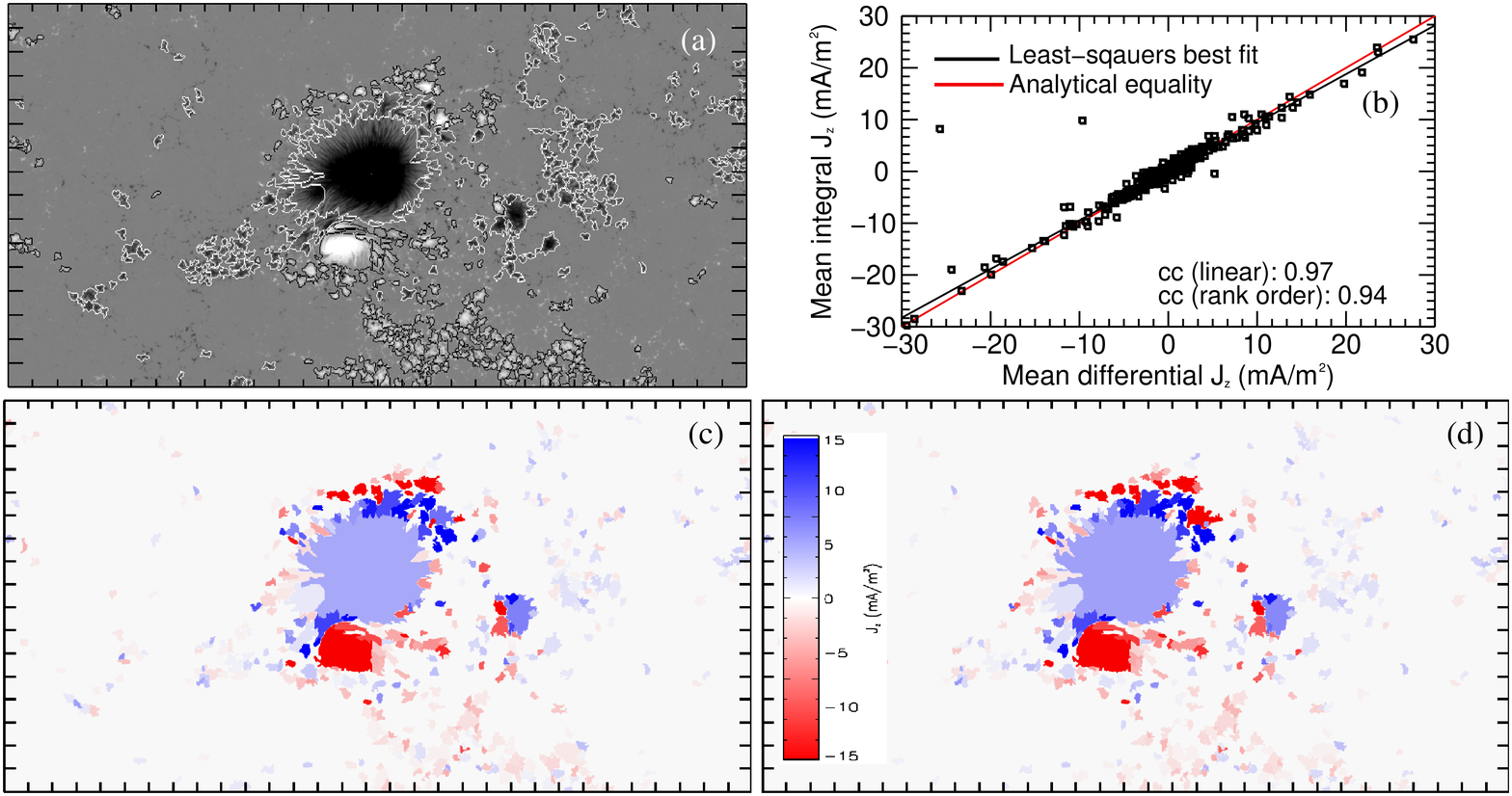}}
\figcaption{Flux and current partitioning of NOAA AR 10930. (a) The
  partitioned vertical magnetic field, saturated at $\pm 2000\;G$. 
  Different partitions are indicated by their enclosing contours.  
  (b) Comparison between the mean vertical electric current densities per
  partition, as calculated using the integral (shown in c) and the
  differential (shown in d) forms of Amp\'{e}re's law. 
  Both current-density maps (c,d) are saturated at $\pm 15 mA/m^2$.
  The linear
  (Pearson) and rank-order (Spearman) correlation coefficients (cc)
  between the two current density estimates are also given. 
  Tic mark separation in all images is 10\arcsec.
\label{ar930p}}
\newpage
%
\centerline{\includegraphics[width=12.cm,height=14.5cm,angle=0]{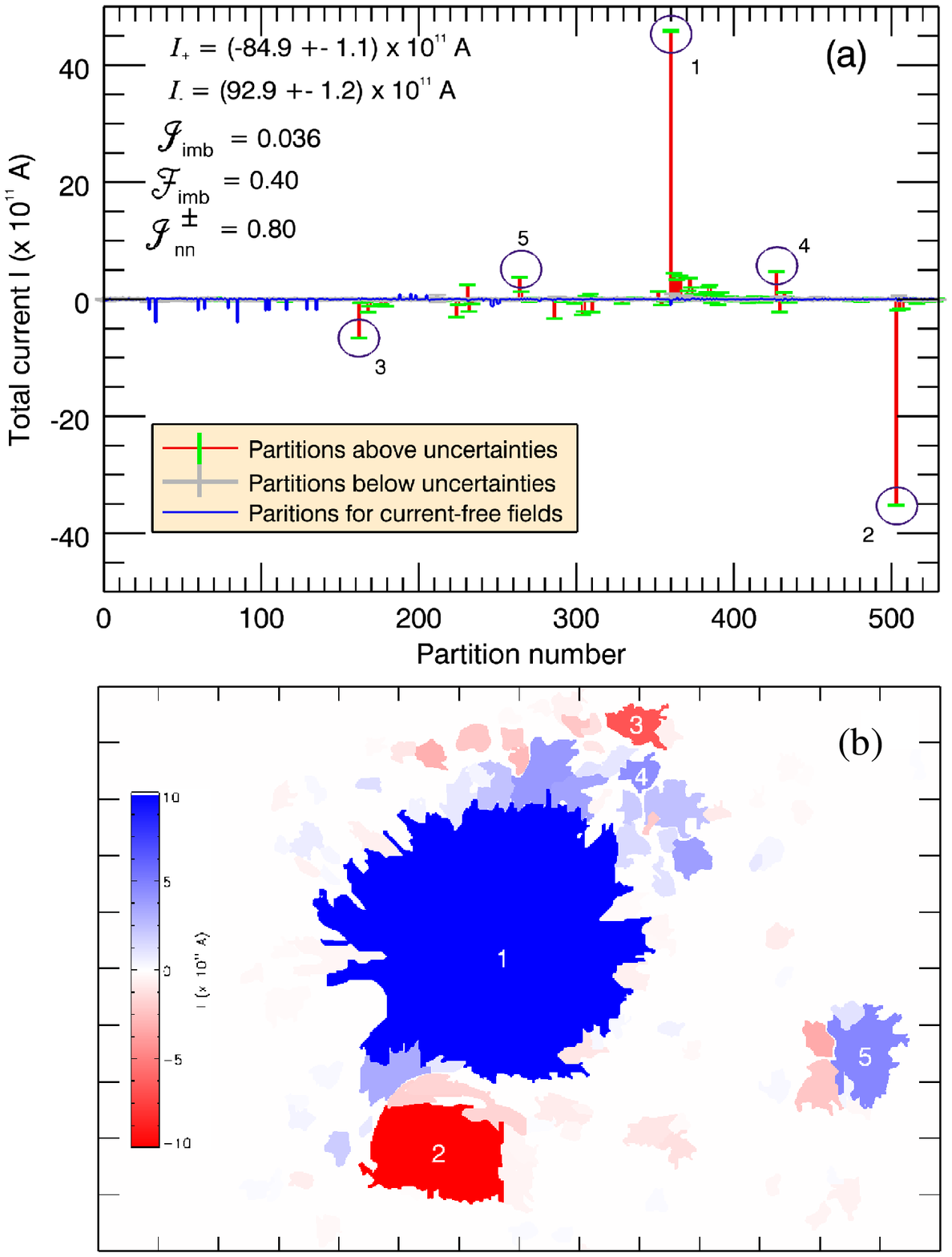}}
\figcaption{Total partition currents in NOAA AR 10930. (a) Total
  currents for each partition. Currents that satisfy the significance
  criteria of equations (\ref{crit}) are shown with red columns and
  green error bars. Five of
  the largest non-neutralized total currents are encircled and labeled (1-5). 
  The total currents $I_{+}$ and $I_{-}$ for positive- and
  negative-polarity partitions, respectively, the total current imbalance
  $\mathcal{I}_{\rm imb}$, the magnetic flux imbalance $\mathcal{F} _{\rm imb}$, and   
  the non-neutrality factor $\mathcal{I}^{\pm}_{\rm nn}$ are also
  given. (b) Detail of the active region showing the spatial 
  distributions of the partitions that satisfy the
  criteria of equations (\ref{crit}).
 Different partitions are indicated by different shades of blue and red, according to their total current. Total currents are saturated at $\pm 10^{12}\;A.$ Partitions 1-5, with total currents
  corresponding to the five currents identified in a, are also
  indicated. Tic mark separation is 10\arcsec.    
\label{ar930cl}}
\newpage
\centerline{\includegraphics[width=16.5cm,height=10.cm,angle=0]{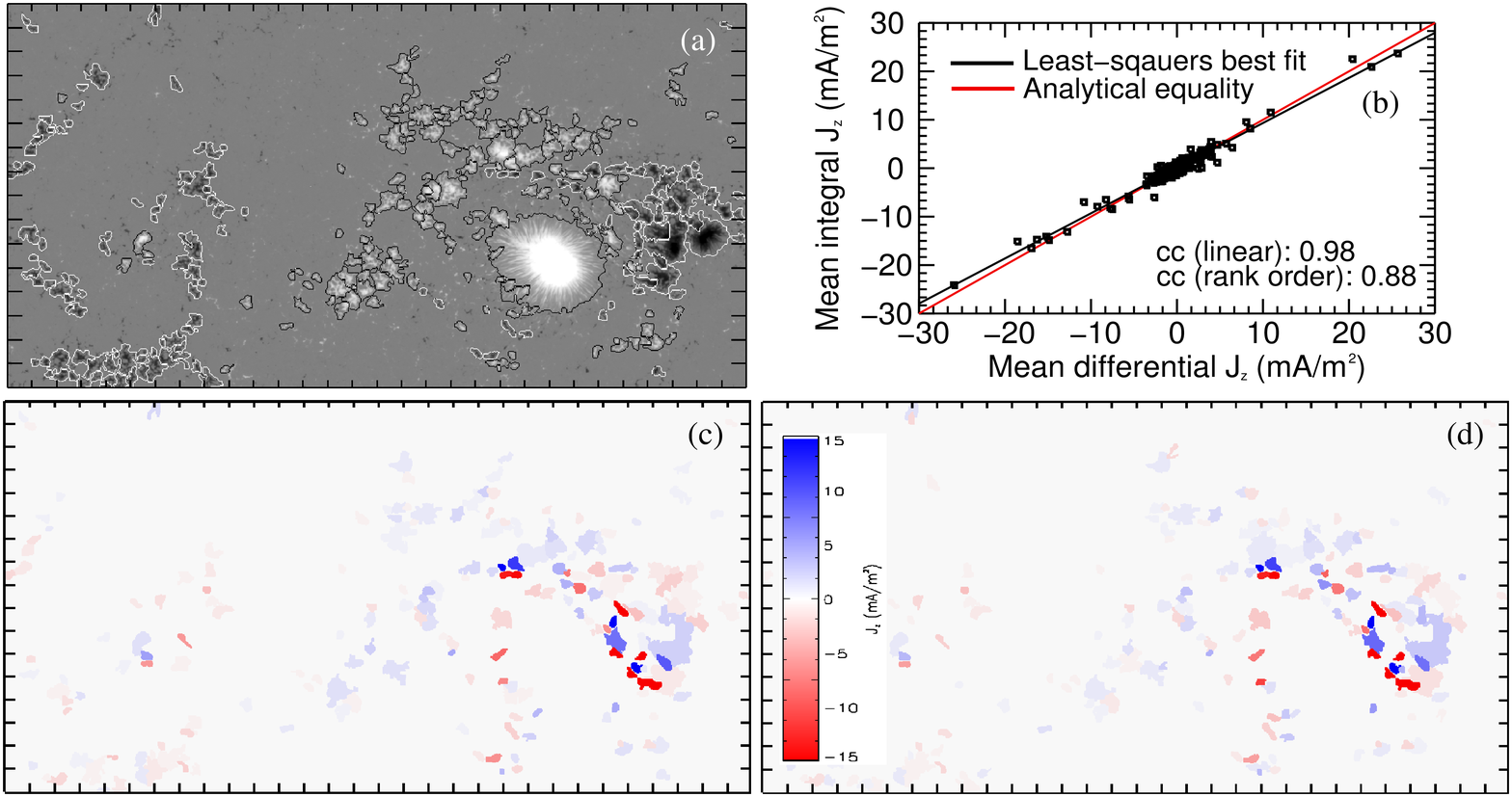}}
\figcaption{Same as Figure \ref{ar930p}, but for NOAA AR 10940.
\label{ar940p}}
\newpage
%
\centerline{\includegraphics[width=12.cm,height=15.cm,angle=0]{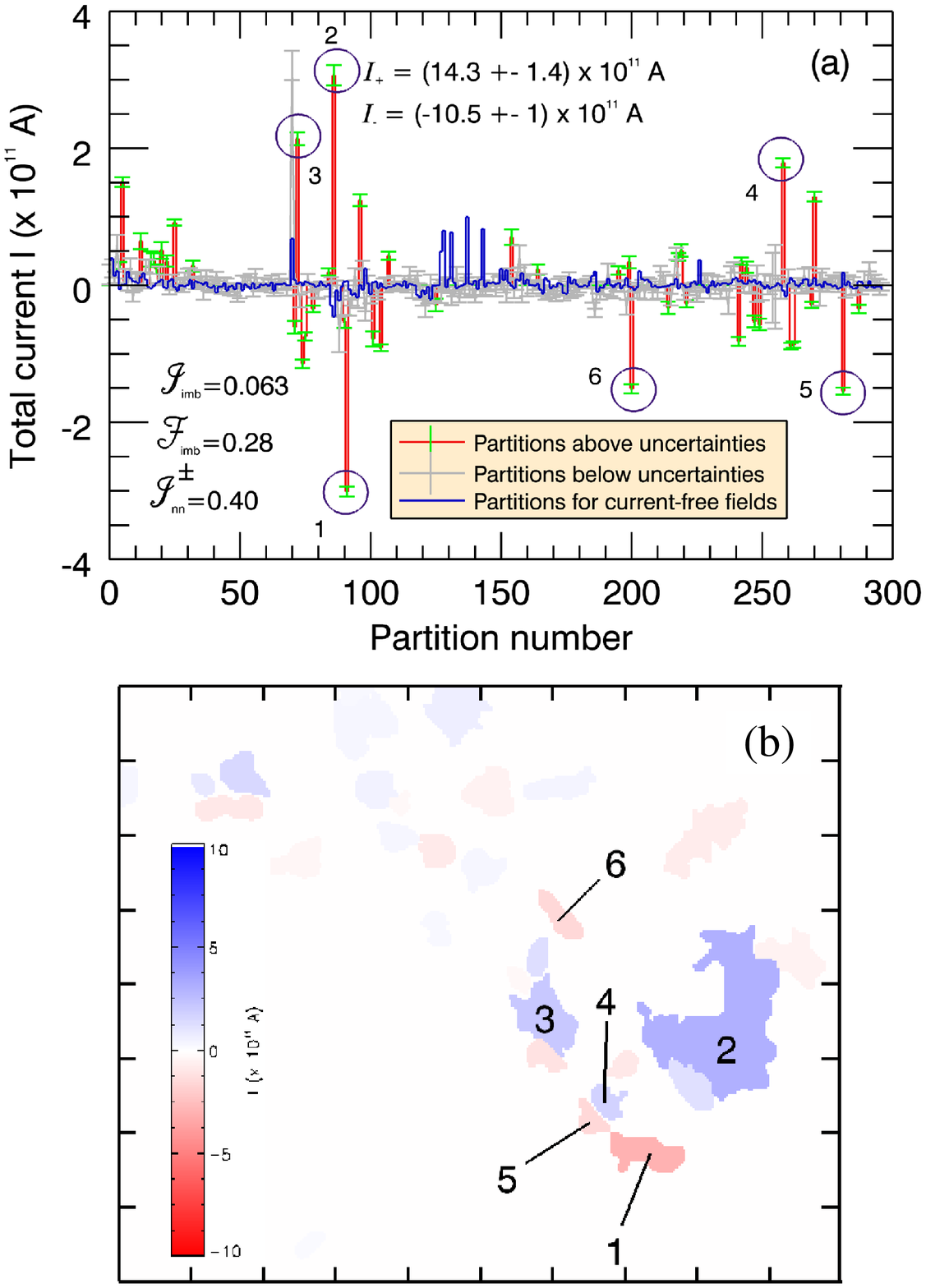}}
\figcaption{Same as Figure \ref{ar930cl}, but for NOAA AR 10940. Six
  partitions with the largest non-neutralized total currents are
  identified here.
\label{ar940cl}}
\newpage
\centerline{\includegraphics[width=12.cm,height=14.cm,angle=0]{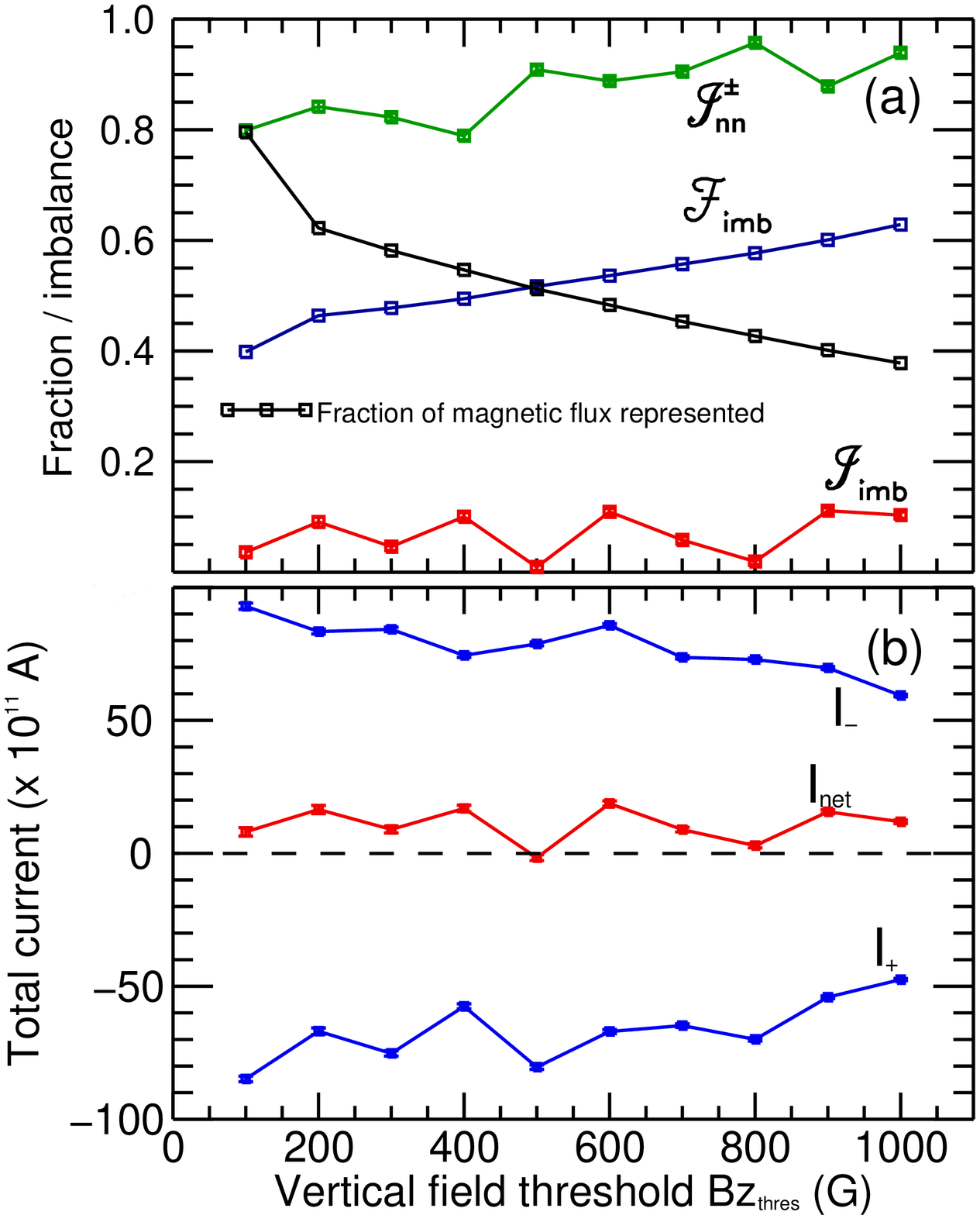}}
\figcaption{Total electric currents in NOAA AR 10930 for different partitioning thresholds 
  $B_{z_{thres}}$. (a) Curves show 
  the fraction of the magnetic flux represented by the partitioning (black), 
  the total current imbalance $\mathcal{I}_{\rm imb}$ (red), 
  the magnetic flux imbalance $\mathcal{F}_{\rm imb}$ (blue), and 
  the current non-neutrality factor
  $\mathcal{I}^{\pm}_{\rm nn}$ (green) for each $B_{z_{thres}}$.
  (b) The total currents $I_{+}$ and $I_{-}$ for the positive and negative
  polarity, respectively (blue curves) and the net current $I_{net}$ (red
  curve). 
\label{ar930_stat}}
\centerline{\includegraphics[width=12.cm,height=14.cm,angle=0]{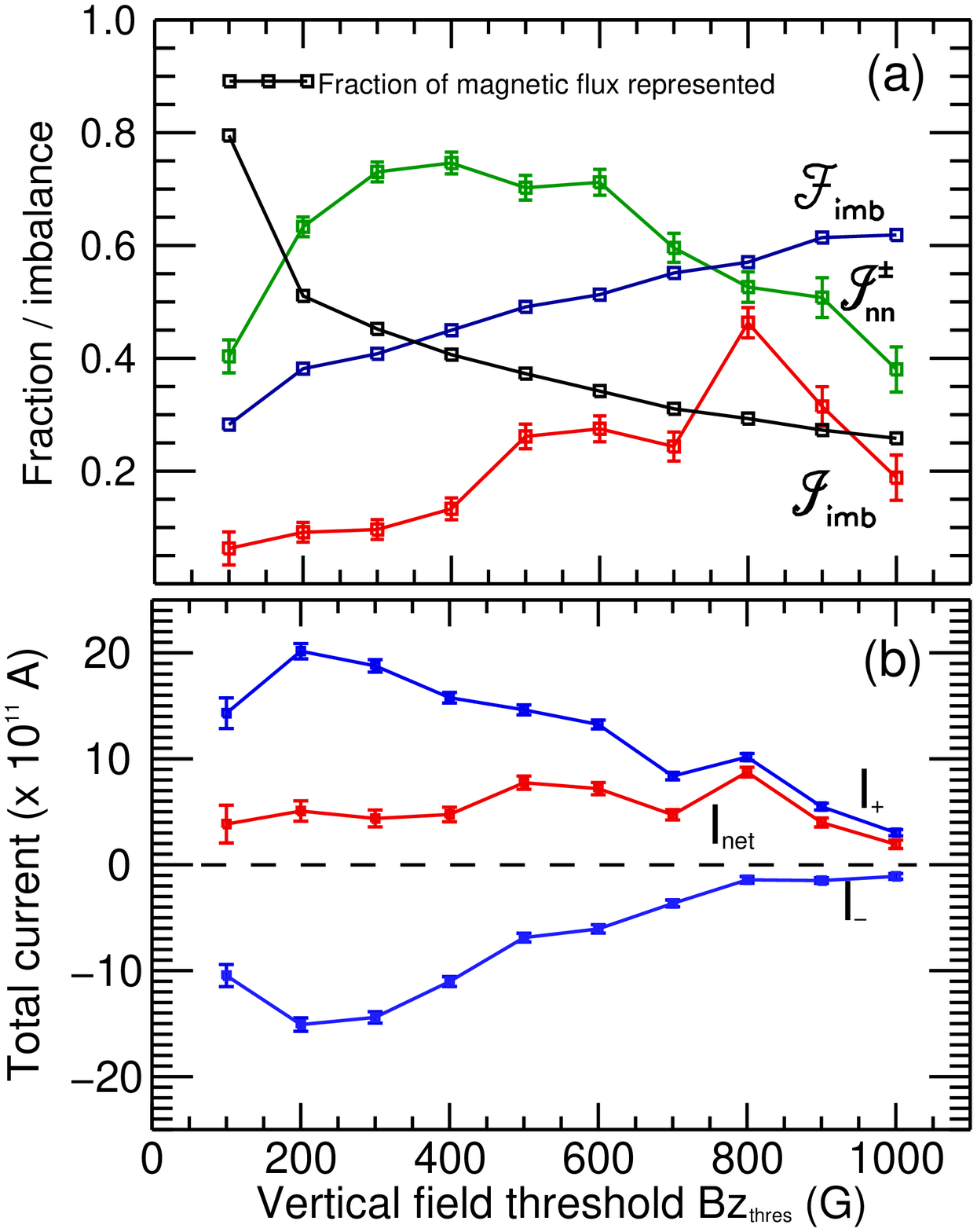}}
\figcaption{Same as Figure \ref{ar930_stat}, but for NOAA AR 10940.
\label{ar940_stat}}
\newpage
\centerline{\includegraphics[width=14.cm,height=7.cm,angle=0]{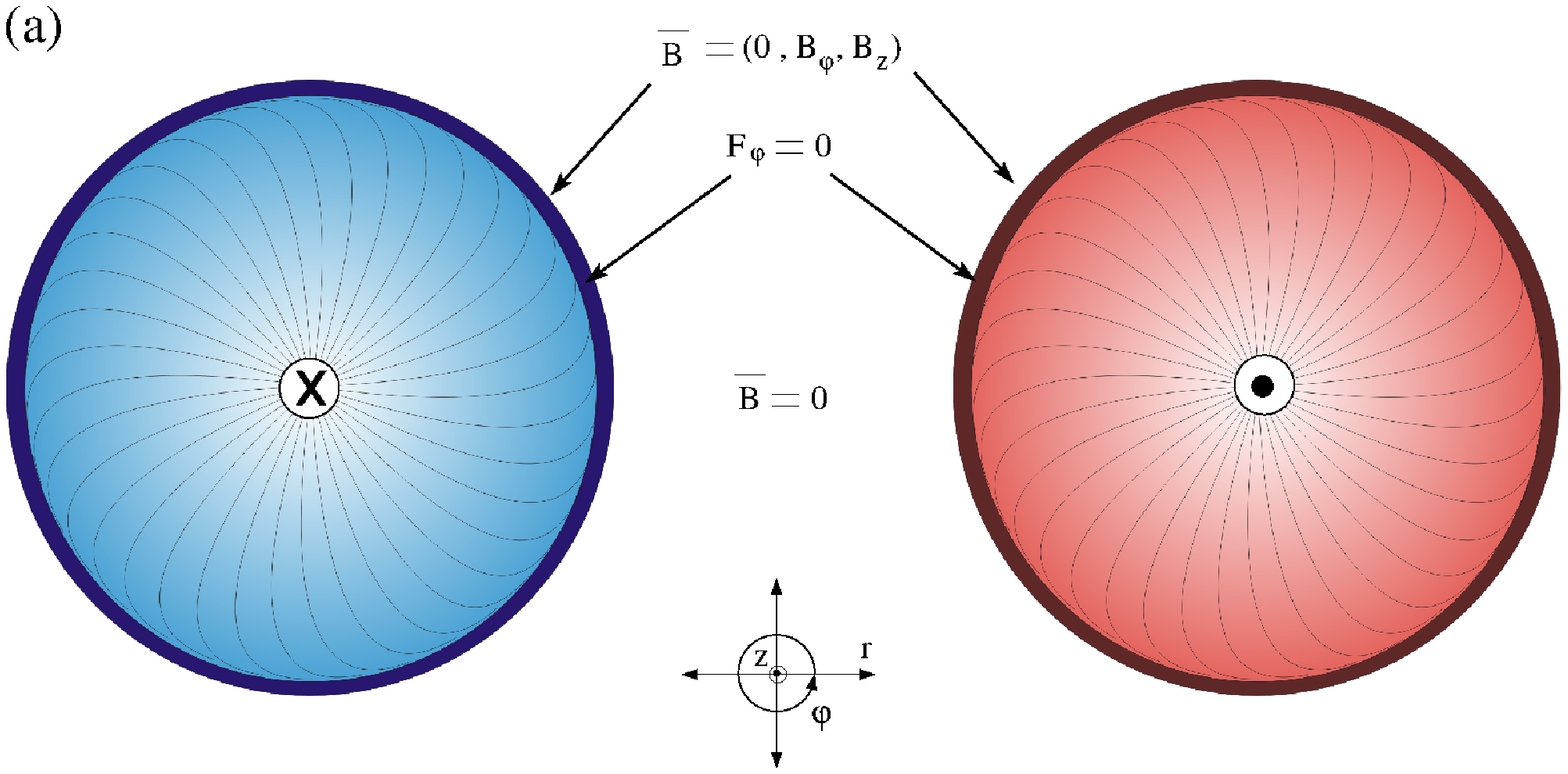}}
\centerline{\includegraphics[width=10.cm,height=8.cm,angle=0]{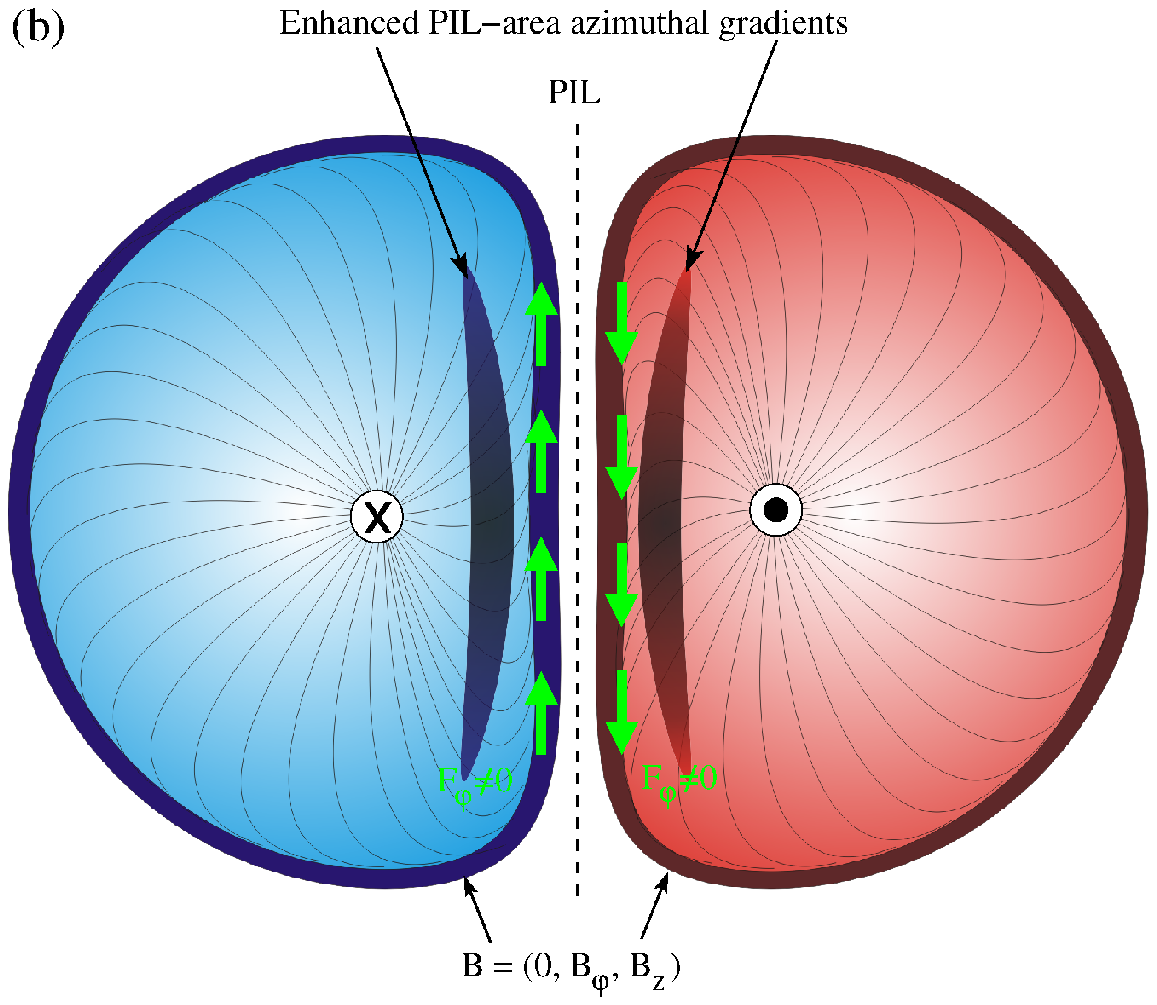}}
\figcaption{Simplified concept for the occurrence of shear-generating Lorentz forces along lower-boundary PILs in solar active regions. The situation is qualitatively similar for different altitudes, along the normal direction $z$, until the field becomes space-filling. (a) In case of non-interacting flux-tube footprints, hence no well-defined PILs, a lack of azimuthal gradients within and on the interface between the flux tube and the field-free space inhibits a shearing (azimuthal) Lorentz force $F_\varphi$.
(b) In case of interacting, deformed flux tubes, inhomogeneous azimuthal gradients are formed (dark shapes), preferentially enhancing sheath currents along the PIL (dashed line) that become non-neutralized within a given polarity and give rise to an azimuthal, shearing Lorentz force $F_\varphi$ in the PIL area. In both panels, the field polarity is denoted at the center of each footprint, while red- and blue-color gradients suggest the gradient of $\mbf{B}$. Spiral curves denote twist in the flux tube, and hence the presence of an azimuthal field $B_\varphi$. Each footprint is enclosed by its sheath currents 
($\sim \nabla B \times \mbf{B}$ - dark red and blue rings) for the flux tube to maintain coherence.  
\label{fl_cartoon}}
%
%
\newpage
\begin{table}
\begin{tabular}{ccccccccc}
\hline
\hline
Pixel size   & $N$ & $N_{\rm nn}$ & $I_{+}$  & $I_{-}$ & $I_{\rm net}$ & $\mathcal{I}_{\rm imb}$ & $\mathcal{I}^{\pm}_{\rm nn}$  & $\mathcal{F}_{\rm imb}$ \\
($arcsec$) &        &                     & ($\times 10^{11}\;A$) &  ($\times 10^{11}\;A$) & ($\times 10^{11}\;A$) & & \\
\hline
0.1585& 531 & 120 & $-84.9 \pm 1.1$ & $92.9 \pm 1.2$ & $8.1 \pm 1.6$ & 0.036  & 0.799  & 0.398\\
0.317 & 449 &  76 & $-86.1 \pm 0.8$ & $89.7 \pm 0.9$   & $3.7 \pm 1.1$ & 0.018 &  0.867 & 0.447\\
0.634 &  65 &  19 & $-58.7 \pm 0.6$ & $70.3 \pm 0.6$ & $11.6 \pm 0.8$ & 0.086&  0.967 & 0.583\\
1.268 &  12 &   6 & $-47.9 \pm 0.4$ & $57.2 \pm 0.6$ & $9.3 \pm 0.7  $ & 0.089&  1.000 & 0.678\\
2.536 &   3 &   2 & $-38.8 \pm 0.6$ & $50.9 \pm 0.6$ & $12.1 \pm 0.8$  & 0.132&  0.978 & 0.665\\
\hline
\hline
\end{tabular}
\caption{Effect of varying spatial resolution on the total
  currents of NOAA AR 10930; $N$ is the total number of partitions,
  $N_{\rm nn}$ is the number of non-neutralized partitions. The pixel size has been 
  modified by binning the original Hinode SOT/SP magnetogram of the AR. Shown are the total currents per polarity ($I_{+}, I_{-}$), the net current ($I_{\rm net}$), the total-current imbalance ($I_{\rm imb}$), the current non-neutrality factor ($\mathcal{I}^{\pm}_{\rm nn}$), and the magnetic-flux imbalance ($\mathcal{F}_{\rm imb}$).}
\label{tb1}
\end{table} 
\begin{table}
\begin{tabular}{ccccccccc}
\hline
\hline
Pixel size   & $N$ & $N_{\rm nn}$ & $I_{+}$  & $I_{-}$ & $I_{\rm net}$ & $\mathcal{I}_{\rm imb}$ & $\mathcal{I}^{\pm}_{\rm nn}$  & $\mathcal{F}_{\rm imb}$ \\
($arcsec$) &        &                     & ($\times 10^{11}\;A$) &  ($\times 10^{11}\;A$) & ($\times 10^{11}\;A$) & & \\
\hline
0.317 & 297 & 45 & $14.3 \pm 1.5$ & $-10.46 \pm 1$ & $3.8 \pm 1.8$ & 0.063 & 0.403 & 0.282\\
0.634 &  50 &  9 & $11.5 \pm 0.5$ & $-7.2 \pm 0.5$ & $4.2 \pm 0.7$ & 0.156 & 0.681 & 0.474\\
1.268 &  10 &  2 & $7.0 \pm 0.6$ & $-1.2 \pm 0.4$ & $5.8 \pm 0.7$ &  0.514 & 0.640 & 0.581\\
2.536 &   3 &  1 & $3.9 \pm 0.5$ & $2.1 \pm 0.5$ & $6.0 \pm 0.7$ & 0.580 & 0.663 & 0.541\\
\hline
\hline
\end{tabular}
\caption{Same as Table \ref{tb1} but for NOAA AR 10940.} 
\label{tb2}
\end{table} 
\end{document}